\documentclass[oneside,a4paper,reqno]{amsart}
\usepackage{amsmath}
\usepackage{amssymb}
\usepackage{xspace}
\usepackage[mathscr]{eucal}
\usepackage{graphics}
\newcommand{\mat}[3]{\ensuremath{
																								\left \langle  \vphantom{#2 #3}   #1   
                        \right|    					\, #2\,   
                        \left|    \vphantom{#2 #1} #3   
                        \right \rangle
                        								}
                     }

\newcommand{\ket}[1]{\ensuremath{		\left| #1 \right> 
																																			  }
																									}

\newcommand{\bra}[1]{\ensuremath{		\left< #1 \right| 
																																			  }
																									}

\newcommand{\overlap}[2]{\ensuremath{ 
																								\left \langle    #1 \vphantom{#2 } \,
                        \right| \left.   #2 \vphantom{#1}
                        \right \rangle
                        									}
                     }
\newcommand{\boverlap}[2]{\ensuremath{ 
																								\bigl \langle #1 \, \bigr| \bigl. #2 \bigr \rangle
                        									}
                     }

\newcommand{\vb}{v_{\mathrm{B}}}

\newcommand{\bve}{\bar{v}_{\mathrm{E}}}

\newcommand{\tcrit}{t_{\mathrm{c}}}
\newcommand{\tpcrit}{\tilde{t}_{\mathrm{c}}}
\newcommand{\tloc}{t_{\mathrm{loc}}}

\newcommand{\amp}{x_{\mathrm{max}}}
\newcommand{\rhor}{\hat{\rho}_{\mathrm{red}}}
\newcommand{\bolt}{k_{\mathrm{B}}}
\newcommand{\deBrog}{\lambda_{\mathrm{B}}}

\DeclareMathOperator{\imaginary}{Im}

\begin{document}\begin{titlepage}
\begin{center}
\bfseries
BOHMIAN TRAJECTORIES POST-DECOHERENCE
\end{center}
\vspace{1 cm}
\begin{center}
D M APPLEBY
\end{center}
\begin{center}
Department of Physics, Queen Mary and
		Westfield College,  Mile End Rd, London E1 4NS, UK
 \end{center}
\vspace{0.5 cm}
\begin{center}
  (E-mail:  D.M.Appleby@qmw.ac.uk)
\end{center}
\vspace{0.75 cm}
\begin{center}
  QMW--PH--99--11
\end{center}
\vspace{1.25 cm}
\begin{center}
\textbf{Abstract}\\
\vspace{0.35 cm}
\parbox{10.5 cm }{ The proposal that the interaction between a macroscopic 
body and its environment plays a crucial role in producing the correct 
classical limit in the Bohm interpretation of quantum mechanics is investigated, in
the context of a model of quantum Brownian motion.  It is well known that one of the
effects of the interaction is to produce an extremely rapid approximate
diagonalisation of the reduced density matrix in the position representation.  This
effect is, by itself, insufficient to produce generically quasi-classical behaviour of
the Bohmian trajectory.  However, it is shown that, if the system particle is
initially in an approximate energy eigenstate, then there is a tendency for the
Bohmian trajectory to become approximately classical on a rather longer time-scale.
The relationship between this phenomenon and the behaviour of the 
Wigner function post-decoherence (as analysed by Halliwell and
Zoupas) is
discussed.  It is also suggested that the phenomenon may be
related to the storage of information about the trajectory in the environment,
and that  it may therefore be a general feature of every situation in which  such
environmental monitoring occurs.
                      }
\end{center}
\end{titlepage}
\title{Bohmian Velocity post-Decoherence}
\section{Introduction}  
\label{sec:  Introduction}
It is
well-known~\cite{Bohm2,Holland1,Holland2,self1} that the trajectories in
the Bohm Interpretation of Quantum Mechanics are often highly
non-classical.  This gives rise to an important problem for the Bohm
interpretation:  namely, the question as to how the interpretation can
account for existence of generically quasi-classical trajectories on the
macroscopic level.  In a previous paper~\cite{self1} we argued that the
Bohm interpretation typically fails to produce the correct classical limit
if the system is isolated.  The purpose of this paper is to argue that the
correct classical limit does emerge once one takes into the account the
effect of the environment.  Our discussion extends the analysis given
in Chapter 8 of the book by Bohm and Hiley~\cite{Bohm2}.

We are particularly interested in the connection between the role
of the environment in the Bohm interpretation, and the phenomenon of
decoherence, which plays a central role in the decoherent histories
approach~\cite{Griff,GellMann2,GellMann1,Omnes}, and in Zurek's existential
interpretation~\cite{Zurek2,Zurek1,Zurek3}.
The mechanism considered by Bohm and Hiley---the scattering of a beam of
 radiation or other particles---is also one of the
mechanisms by which decoherence is produced~\cite{Omnes,Joos,Giulini}. 
This
has suggested to some authors~\cite{Dickson,Zeh} that 
the process by which the Bohmian trajectory becomes quasi-classical is
closely related to the phenomenon of  decoherence. The suggestion is
certainly plausible.   However,  it is not entirely clear, just from the
argument given by Bohm and Hiley, that the suggestion is actually correct.
At the macroscopic level decoherence is an ubiquitous phenomenon, which
can be produced by a wide variety of different mechanisms.  By contrast,
Bohm and Hiley only consider the particular case of a scattering process. 
One would like to know whether other kinds of interaction between a
macroscopic body and its environment also have the effect of causing the
Bohmian trajectory to become approximately classical.  More generally, one
might ask whether every process which causes decoherence also causes the
Bohmian trajectory to become approximately classical, or whether it is
only some of them.  These are the questions which motivated our
investigation.

We will focus on the models of quantum Brownian motion which have
been discussed by 
Caldeira and Leggett, Hu, Paz and Zhang, and many
others~\cite{Caldeira,Grab,Hu,Hu2}.  These models have played an
important role in studies of decoherence, and are therefore a
natural starting point for an investigation into the role of decoherence in
the Bohm interpretation.  We accordingly consider a system particle, with
position
$\hat{x}$ and momentum $\hat{p}$, interacting with a heat bath consisting
of $N$ other particles with positions $\hat{x}_1, \dots, \hat{x}_N$ and 
momenta $\hat{p}_1, \dots, \hat{p}_N$.  The Hamiltonian is
\begin{equation}
\hat{H}  =
\left(\frac{1}{2 m} \hat{p}^2 + \frac{1}{2} m \omega_{0}^2 \hat{x}^2 \right)
+
\sum_{r=1}^{N}
\left(\frac{1}{2 m_{r}}
\hat{p}_{r}^2 +
\frac{1}{2} m_{r}
                       \omega_{r}^2  \hat{x}_{r}^2\right)+
                       \sum_{r=1}^{N} \kappa_{r} \hat{x}
\hat{x}_{r}
\label{eq:  BrownHamDef}
\end{equation}
Here $\omega_{0}$ denotes the bare frequency of the  system
particle.  The
renormalised frequency will be denoted $\omega$.

The model is characterized by the spectral density
\begin{equation}
  I(\omega') = 
  \sum_{r=1}^{N} \frac{\kappa_{r}^{2}}{2 m_{r} \omega_{r}}
   \delta(\omega'-\omega_{r})
\label{eq:  IomegDef}
\end{equation}
Taking $I\propto\omega'$ (for $\omega'<$ the cut-off frequency) gives the
Caldeira-Leggett model~\cite{Caldeira}. If one leaves $I$ arbitrary one
obtains the  general class of master  equations derived by Hu, Paz and
Zhang~\cite{Hu}.

At $t=0$ the heat bath is taken to be in the thermal state with 
density matrix
\begin{equation}
  \hat{\rho}_{\mathrm{bath}}
= \mathscr{N} \exp \left( - \frac{\hat{H}_{\mathrm{bath}}}{
  \bolt T}\right)
\label{eq:  RhoBatDef}
\end{equation}
where
\begin{equation*}
  \hat{H}_{\mathrm{bath}} = \sum_{r=1}^{N}
\left(\frac{1}{2 m_{r}}
\hat{p}_{r}^2 +
\frac{1}{2} m_{r}
                       \omega_{r}^2  \hat{x}_{r}^2\right)
\end{equation*}
and $\mathscr{N}$ is a constant.  We assume  that at $t=0$ system$+$heat bath are
in the product state $\ket{\psi_{\mathrm{sys}}} \bra{\psi_{\mathrm{sys}}}
\otimes \hat{\rho}_{\mathrm{bath}}$.  At
$t\ge 0$ the density matrix describing system$+$heat bath will consequently be
\begin{equation*}
  \hat{\rho}(t) = e^{-i t \hat{H}/\hbar}
    \Bigl(  \ket{\psi_{\mathrm{sys}}} \bra{\psi_{\mathrm{sys}}}
  \otimes  \hat{\rho}_{\mathrm{bath}} \Bigr)
  e^{i t \hat{H}/\hbar}
\end{equation*}
In the conventional approach one now integrates
out the environmental degrees of freedom, and focusses on the behaviour
of the
reduced density matrix.  Unfortunately matters are not so simple in 
the Bohm interpretation.

In the Bohm interpretation a mixed state such as
$\hat{\rho}_{\mathrm{bath}}$ is taken to describe an ensemble of pure states
\begin{equation}
  \hat{\rho}_{\mathrm{bath}} 
= \sum_{\alpha} \rho_{\alpha} \ket{\phi_{\alpha}}\bra{\phi_{\alpha}}
\label{eq:  BathEnsemble}
\end{equation}
This way of writing the density matrix is not simply a mathematical
device, as in the conventional approach. Rather, one takes it that at
$t=0$
the heat bath actually is in  one of the pure states in the ensemble, 
with $\rho_{\alpha}$ being the probability that it is in the state 
$\ket{\phi_{\alpha}}$.   The  problem we then  face is, that the density
matrix does not uniquely determine the ensemble, and that in the Bohm
interpretation it makes a difference  which ensemble we choose 
(for a  classification of
the set of all discrete ensembles corresponding to a given density  matrix
see Hughston
\emph{et al}~\cite{Hugh}).  We discuss
this point further in Section~\ref{sec:  QuantBrown}.

Suppose that a particular ensemble has been chosen, and suppose that at
$t=0$ the heat bath is in the pure state
$\ket{\phi_{\alpha}}$.  Then at $t>0$ system$+$heat bath will be in the 
pure state
\begin{equation}
  \ket{\Psi_{\alpha}(t)} = e^{- i t \hat{H}/\hbar}
\ket{\psi_{\mathrm{sys}}}\otimes\ket{\phi_{\alpha}}
\label{eq:  SysHtBth}
\end{equation}
and the Bohmian velocity of the system particle will be given by
\begin{equation}
  \vb^{(\alpha)} (t,x,x_1,\dots,x_N)
= 
\frac{\hbar\imaginary 
       \bigl(\overlap{\Psi_{\alpha}(t)}{x,x_1,\dots,x_N}
             \frac{\partial}{\partial x}
             \overlap{x,x_1,\dots,x_N}{\Psi_{\alpha}(t)}
       \bigr)
     }{m \bigl|  \overlap{x,x_1,\dots,x_N}{\Psi_{\alpha}(t)}\bigr|^2
     }
\label{eq:  BVelDef}
\end{equation}
We see that the Bohmian velocity depends, not only on $x$, but also on 
$x_1, \dots, x_N$, as well as the index $\alpha$.  The reduced density
matrix clearly does not provide enough information to calculate this
function.  Consequently, the problem of determining the effect of the
interaction with the heat bath is significantly more difficult in the Bohm
interpretation than it is in the conventional approach.  

Nevertheless, although the reduced density matrix does not provide us with
complete information regarding the Bohmian velocity of the system particle,
it does tell us something.  To see this, consider the effect of averaging
$\vb^{(\alpha)}$ over all the possible values of  $x_1, \dots,
x_N$, and of the index $\alpha$:
\begin{equation*}
 \bve(t,x)=\frac{\sum_{\alpha} \rho_{\alpha} \int dx_{1} \dots dx_{N} \,
    \bigl|\boverlap{x,x_1,\dots,x_N}{\Psi_{\alpha}(t)}\bigr|^2 \,
    \vb^{(\alpha)}(t,x,x_{1},\dots,x_{N}) }{
    \sum_{\alpha} \rho_\alpha \int dx_1 \dots dx_N \,
    \bigl|\boverlap{x,x_1,\dots,x_N}{\Psi_{\alpha}(t)}\bigr|^2}
\end{equation*}
We will refer to $\bve$ as the ensemble-averaged velocity.
The reduced
density matrix elements are given by
\begin{multline*}
 \mat{x}{ \rhor(t)}{x'} \\
= \sum_{\alpha} \rho_{\alpha} \int dx_1 \dots dx_N \,
   \boverlap{x,x_1,\dots,x_N}{\Psi_{\alpha}(t)}
   \boverlap{\Psi_{\alpha}(t)}{x',x_1,\dots,x_N}
\end{multline*}
It is then straightforward to infer
\begin{equation}
  \bve(t,x) =
  \frac{\hbar}{m}
  \frac{\imaginary \left(\left. \frac{\partial}{\partial x}
     \mat{x}{\rhor(t)}{x'} \right|_{x'=x}\right)}{
     \mat{x}{\rhor(t)}{x}}
\label{eq:  bveTermsRhoRed}
\end{equation}
from which we see that the reduced density matrix does provide enough
information to calculate $\bve$.

In this paper we will investigate the behaviour of $\bve$ and 
$\vb^{(\alpha)}$ as  functions of time for the case when the initial
system state is an approximate energy eigenstate, of the form~\cite{self1}
\begin{equation}
  \ket{\psi_{\mathrm{sys}}}  = \sum_{r=\bar{n}-\frac{\Delta
n}{2}}^{\bar{n}+\frac{\Delta n}{2}}
   c_r \ket{\bar{n} +r}
\label{eq:  ProxESteDef}
\end{equation} 
In this expression  $\ket{n}$ denotes the $n^{\mathrm{th}}$ 
eigenstate of the isolated  system particle Hamiltonian:
\begin{equation*}
  \hat{H}_{\mathrm{sys}}\ket{n}= E_{n} \ket{n}
\end{equation*}
where $\hat{H}_{\mathrm{sys}} = \hat{p}^2/(2m)+m \omega^2 \hat{x}^2/2$
and 
$E_{n}=\left(n+1/2\right)\hbar \omega$.  We assume that
$\bar{n}
\gg 1$ (so that the state is highly excited), and
$\Delta n
\ll
\bar{n}$ (so that the energy distribution is  sharply peaked
about the mean).
Classically one would therefore expect the  particle to be following
a well-defined orbit with energy close to $E_{\bar{n}}$ and amplitude
close to $\amp = \left(2 E_{\bar{n}}/(m \omega^2)\right)^{1/2}$.  In
particular, when the particle is located at  $x$, one would
classically expect
its velocity to be close to  $\pm
p_{\mathrm{cl}}(x)/m$, where
\begin{equation*}
  p_{\mathrm{cl}}(x)  = m \omega
\left(\amp^2-x^2\right)^{\frac{1}{2}}
\end{equation*} 
On the other hand it was shown in ref.~\cite{self1} that, assuming the
system to be isolated, there is only a high probability of this being true
of the Bohmian velocity at all stages of the motion in the very special case 
for which
$\ket{\psi_{\mathrm{sys}}}$ is a  narrowly localized wave packet.
In the following we will show that the effect of the interaction with the
heat bath is to make the distribution of Bohmian velocities eventually 
become  approximately classical, whether or not this is true
initially. 

We begin,
in  Sections~\ref{sec:  ProxDiag} and~\ref{sec:  PostDecoherence},
by considering the
behaviour of the ensemble-averaged velocity
$\bve$.  
The feature of the interaction with the environment which has probably
attracted most attention is the tendency of the reduced density matrix
to become approximately diagonal in the position representation.  In 
Section~\ref{sec:  ProxDiag} we show that this phenomenon is, \emph{by
itself}, insufficient to produce approximately classical behaviour of the
Bohmian trajectory.
However, the interaction has other important effects, apart from the
approximate diagonalisation of $\rhor$.  In particular, Halliwell and
Zoupas~\cite{HalliB} have shown that, in the case of the Caldeira-Leggett
model, the Wigner function  becomes non-negative after a sufficient elapse
of time.  In Section~\ref{sec:  PostDecoherence} we show that, as a
consequence of this effect,
$\bve$ comes to lie approximately within the classical range 
\begin{equation}
-p_{\mathrm{cl}}(x)/m \le \bve(x) \le p_{\mathrm{cl}}(x)/m 
\label{eq:  bveClassCond}
\end{equation}
We also derive conditions for this  to occur in the case of 
other models of the type defined by 
Eq.~(\ref{eq:  BrownHamDef}) (and, in fact, for a number of models
which are not of this type).

Inequalities~(\ref{eq:  bveClassCond}) represent a necessary condition for
the Bohmian trajectory to be approximately classical.  However, they are
clearly not sufficient.  In Section~\ref{sec:  QuantBrown} we accordingly 
calculate the function $\vb^{(\alpha)}(t,x,x_1,\dots,x_N)$ on the assumption
that the ensemble described by $\hat{\rho}_{\mathrm{bath}}$ consists of
coherent states
[see remarks following Eq.~(\ref{eq:  BathEnsemble})].
Our calculation is based on Halliwell and Yu's alternative 
derivation~\cite{HalliC} of
the Hu-Paz-Zhang master equation (also see Anglin and
Habib~\cite{AnglinB}), which has the advantage (from our point of view)
that, unlike the usual path integral methods, it allows us  explicitly to
keep track of the heat bath degrees of freedom.  We show that in the case
of the Caldeira-Leggett model, for sufficiently large values of $t$, there
is a high probability that $\vb^{\alpha}$ will be close to one of the 
classical values $\pm p_{\mathrm{cl}}(t,x,x_1,\dots,x_N)$.
We also derive conditions for this  to occur in the case of 
other models of the type defined by 
Eq.~(\ref{eq:  BrownHamDef})

Finally, in the conclusion, we discuss the bearing that these results have
on the questions which provided the original motivation for this
investigation, and we suggest some directions for further enquiry.

\section{Effect of Approximately Diagonalising $\rhor$ in the
$x$-representation.}
\label{sec:  ProxDiag}
One of the most striking effects of the interaction between a macroscopic
body and its environment is that the reduced density matrix tends
rapidly to become approximately diagonal in the position representation.
We begin by showing that this effect is not, by itself, sufficient to cause
the  Bohmian trajectory to become quasi-classical.  
 
 The point is most
conveniently illustrated in the context of the Caldeira-Leggett model, for
which the master equation is~\cite{Caldeira}
\begin{equation}
i \hbar \frac{\partial}{\partial t}\rhor
= \bigl[ \hat{H}_{\mathrm{sys}}, \rhor\bigr] 
  + \gamma \bigl[ \hat{x},
\bigl\{\hat{p},\rhor\bigr\}\bigr]
 - \frac{2 i m \gamma \bolt T}{\hbar} 
\bigl[\hat{x},\bigl[ \hat{x},
\rhor\bigr]\bigr]
\label{eq:  CLMaster}
\end{equation}
where $\hat{H}_{\mathrm{sys}}$ is the renormalised system particle
Hamiltonian (which, for the purposes of this section, need not be assumed
to be of oscillator form), and where
$\{ .,.\}$ denotes an anti-commutator.   It should be noted that this
equation is not exact, and that it does in fact violate the positivity of
$\rhor$ over very short times~\cite{Diosi1,Diosi2}.  However, it 
provides a good approximation over somewhat longer times.

Under conditions where the last
term on the right hand side of Eq.~(\ref{eq:  CLMaster})
dominates, and provided that
$t$ sufficiently small [but not so small as to render the approximation
of Eq.~(\ref{eq:  CLMaster}) invalid], one
has~\cite{Zurek1,Joos,Giulini}
\begin{equation}
  \mat{x}{\rhor(t)}{x'}
\approx \exp\left[- \Lambda t (x-x')^2\right]
\mat{x}{\rhor(0)}{x'}
\label{eq:  quasiXdiag}
\end{equation}
where $\Lambda = (2 m \gamma \bolt T)/\hbar^2$ is the
localization rate.
In the case of a macroscopic object $\Lambda$ is typically
very large~\cite{Joos,Tegmark}, even when the interaction with
the environment is comparatively weak.  Eq.~(\ref{eq: 
quasiXdiag}) consequently plays an important role in  attempts
to explain the emergence of an effectively classical
statistics of ``facts''~\cite{Kupsch} from an underlying
theory which is fully quantum
mechanical~\cite{
GellMann2,GellMann1,Omnes,Zurek2,Zurek1,Joos,Giulini}.

Substituting the expression given by Eq.~(\ref{eq:  quasiXdiag}) into 
Eq.~(\ref{eq:  bveTermsRhoRed}) we find
\begin{equation}
  \bve (t,x) \approx \bve (0,x)
\label{eq:  bveForRhoDiag}
\end{equation}
However, at $t=0$ system$+$environment are in the product state
$\ket{\psi_{\mathrm{sys}}}\bra{\psi_{\mathrm{sys}}}
\otimes \hat{\rho}_{\mathrm{bath}}$, which
means that
$\bve(0,x)$ coincides with the \emph{actual} Bohmian velocity of the system particle
at $t=0$,
\begin{equation*}
 \bve (0,x)
= \frac{\hbar}{m} 
   \frac{\imaginary
\Bigl(\overlap{\psi_{\mathrm{sys}}}{x}
\frac{\partial}{\partial
x}\overlap{x}{\psi_{\mathrm{sys}}}\Bigr) }{
\bigl|\overlap{x}{\psi_{\mathrm{sys}}}\bigr|^2}
\end{equation*}
It was shown in ref.~\cite{self1} that, for many choices of
$\ket{\psi_{\mathrm{sys}}}$, this quantity tends to take values greatly in 
excess of the classical speed.  In view of Eq.~(\ref{eq:  bveForRhoDiag})
the same must be true of $\bve(t,x)$.  It follows, that the
approximate diagonalisation of $\rhor$ in the $x$-representation 
is not, by itself,
sufficient to produce generically quasi-classical behaviour of the
Bohmian trajectory.

\section{The Behaviour of $\bve$ at Later Times}
\label{sec:  PostDecoherence}
The approximation of Eq.~(\ref{eq:  quasiXdiag}) is only valid for sufficiently
small values of $t$.   We now 
want to investigate the behaviour of the Bohmian velocity over longer time-scales, 
and for other models of Brownian motion, apart from the Caldeira-Leggett model. 
We  will consider the function
$\bve(t,x)$ in this Section, and the function
$\vb^{(\alpha)}(t,x,x_1,\dots,x_N)$ in Section~\ref{sec:  QuantBrown}. 

The advantage of considering $\bve(t,x)$ is that in order to calculate it
one only needs to know the master equation.  We will consider master
equations of the form
\begin{multline}
  i \hbar \frac{\partial}{\partial t}
  \rhor
=
  \frac{1}{2} \bigl[\left(h_{1}(t) \hat{x}^2
+h_{2}(t)\hat{p}^2 + h_{3}(t)
\left\{\hat{x},\hat{p}\right\}\right),\rhor\bigr]
+\gamma(t) \bigl[\hat{x},\bigl\{\hat{p},\rhor\bigr\}\bigr] \\
-\frac{i}{\hbar} \Bigl(
J_{11}(t)\bigl[\hat{p},\bigl[\hat{p},\rhor\bigr]\bigr]
-2 J_{1 2}(t) \bigl[\hat{x},\bigl[\hat{p},\rhor\bigr]\bigr]
+ J_{22}(t) \bigl[\hat{x},\bigl[\hat{x},\rhor\bigr]\bigr]
\Bigr)
\label{eq:  GenMaster}
\end{multline}
for which the right-hand side is quadratic in $\hat{x}$ and $\hat{p}$.
This class includes  equations of the  Hu-Paz-Zhang~\cite{Hu} type, 
corresponding to
Brownian motion models of the kind defined by Eq.(\ref{eq:  BrownHamDef}).  It also
includes those equations of the Lindblad form~\cite{Lindblad} for which the
right-hand side is quadratic in $\hat{x}$ and $\hat{p}$. In particular, it
includes the equation discussed by Di\'{o}si~\cite{Diosi1,Diosi2}.

It is most convenient to work 
in terms of the reduced Wigner function, 
\begin{equation}
  W_{\mathrm{red}}(t,x,p) = \frac{1}{h} \int dy \,
     \exp\left( \frac{i}{\hbar} p y\right)
     \mat{x-\frac{y}{2}}{\rhor(t)}{x+\frac{y}{2}}
\label{eq:  RedWigDefA}
\end{equation}
Expressing 
Eq.~(\ref{eq:  GenMaster}) in terms of $W_{\mathrm{red}}$ we find
\begin{equation}
  \frac{\partial}{\partial t} W_{\mathrm{red}}(t,\eta)
= \sum_{r,s=1}^{2} \left(
  K_{r s}(t) \frac{\partial}{\partial \eta_r}
  \bigl( \eta_s W_{\mathrm{red}}(t,\eta)\bigr)
  +J_{r s}(t) \frac{\partial^2}{\partial \eta_r \partial
\eta_s} 
  W_{\mathrm{red}}(t,\eta)
\right)
\label{eq:  WigEqMot}
\end{equation}
where
\begin{align*}
  \boldsymbol{\eta} & = \begin{pmatrix} x \\ p \end{pmatrix} \\
  \mathbf{K}(t) &  = \begin{pmatrix} - h_3 (t) & - h_2 (t) \\
                         h_1 (t) & 2 \gamma(t) + h_{3}(t)
          \end{pmatrix} \\
 \mathbf{J}(t) &  = \begin{pmatrix} J_{11}(t) & J_{1 2} (t)\\
                         J_{12}(t) & J_{22} (t)
         \end{pmatrix}
\end{align*}
It is  straightforward to verify that the solution to 
Eq.~(\ref{eq:  WigEqMot}) may
be written~\cite{Giulini,Chandra,HalliA}
\begin{multline}
W_{\mathrm{red}}(t,\boldsymbol{\eta})
= \frac{1}{\pi \det \mathbf{A}(t) \sqrt{\det \mathbf{M}(t)}} \\ \times
\int d^{2} \boldsymbol{\eta}' \,
\exp\left[-\left(\boldsymbol{\eta}'-\mathbf{A}^{-1}(t)
\boldsymbol{\eta}\right)^{\mathrm{T}}\mathbf{M}^{-1}(t)
     \left(\boldsymbol{\eta}'-\mathbf{A}^{-1}(t)
\boldsymbol{\eta}\right)\right] W_{\mathrm{red}}(0,\boldsymbol{\eta}')
\label{eq:  RedWigProp}
\end{multline}
where the matrices $\mathbf{A}$ and $\mathbf{M}$ are defined by the
equations
\begin{align}
  \frac{\partial}{\partial t} \mathbf{A} & = - \mathbf{K} \mathbf{A} 
\label{eq:  ADefEq}\\
  \frac{\partial}{\partial t} \mathbf{M} & =
  4 \mathbf{A}^{-1} \mathbf{J} \left(\mathbf{A}^{-1}\right)^{\mathrm{T}}
\label{eq:  MDefEq}
\end{align}
(superscript ``T'' signifying ``transpose'') subject to the
initial conditions
\begin{equation*}
  \mathbf{A}(0)=1 \hspace{0.5 in} \mathbf{M}(0)=0
\end{equation*}

Before proceeding further it will be useful to relate this 
equation to the discussion in the last section.
Specialising to the case of the Caldeira-Leggett master equation,
Eq.~(\ref{eq:  CLMaster}), with 
$\hat{H}$  taking the oscillator form $\hat{p}^2/(2 m) + m \omega^2 \hat{x}^2/2$, one
has
\begin{align*}
  \mathbf{K} & = \begin{pmatrix} 0 & - \frac{1}{m} \\
                      m \omega^2 & 2 \gamma
      \end{pmatrix} \\
 \mathbf{J} & = \begin{pmatrix} 0 & 0 \\ 0 & D\end{pmatrix}
\end{align*}
where $D=2 m \gamma \bolt T$. 
If
$\omega t, \gamma t \ll 1$ 
Eqs.~(\ref{eq:  ADefEq}) and~(\ref{eq:  MDefEq}) then imply
\begin{align}
  \mathbf{A} & \approx \begin{pmatrix} 1 & -\frac{t}{m}\\
                              m \omega^2 t & 1
              \end{pmatrix} \notag\\
  \mathbf{M} & \approx 4 D t
  \begin{pmatrix} \frac{t^2}{3 m^2} & - \frac{t}{2 m} \\
                    - \frac{t}{2 m} & 1
  \end{pmatrix} 
\label{eq:  MforCL}
\end{align}
so that~\cite{HalliA}
\begin{multline}
  W_{\mathrm{red}}(t,x,p) \approx
\frac{\sqrt{3} m}{2 \pi D t^2}
\int dx' dp' \exp\left[ - \frac{3 m^2}{D t^3}
\bigl( x'-x\bigr)^2  \right. \\
\left. - \frac{3 m}{D t^2}
\bigl( x'-x\bigr) \bigl( p'-p\bigr)
-\frac{1}{D t} \bigl( p'-p\bigr)^2 \right]
W_{\mathrm{red}}(0,x',p')
\label{eq:  CLWigProx}
\end{multline}
where we have made the further approximation $\mathbf{A}^{-1}(t)\approx
\mathbf{1}$.
In this  expression the width of
the Gaussian convolution in the $p$-direction is $\propto
t^{\frac{1}{2}}$, whereas the width in the $x$-direction is
$\propto t^{\frac{3}{2}}$.  It follows that, if $t$ is
sufficiently small, there will be a significant degree of
smoothing in the
$p$-direction, but no significant smoothing in the
$x$-direction.  To be specific, suppose that the initial system state 
$\ket{\psi_{\mathrm{sys}}}$ is of the form specified by
Eq.~(\ref{eq:  ProxESteDef}), and suppose that $t\ll \tcrit$,
where
\begin{equation}
  \tcrit = \left( \frac{3 m^2
\deBrog^{2}}{D}\right)^{\frac{1}{3}}
\label{eq:  tcDef}
\end{equation}
where $\deBrog=\hbar/(m \omega \amp)$ is the minimum value of the de Broglie
wavelength.  In that case $W_{\mathrm{red}}$ will be nearly constant over
the width of the Gaussian in the $x$-direction, and we can 
approximately write
{\allowdisplaybreaks
\begin{align*}
  W_{\mathrm{red}}(t,x,p) & \approx
  \frac{\sqrt{3} m}{2 \pi D t^2}
  \int dp' \,
  \Biggl( \int dx' \, \exp\left[- \frac{3 m^2}{D t^3}
                        \bigl(x'-x\bigr)^2 \right.
  \Biggr. \\ &\hspace{0.17 in} \Biggl. \left.
                    -\frac{3 m}{D t^2} 
                 \bigr(x'-x\bigl)\bigr(p'-p\bigl)
                \right]
  \Biggr) \exp\left[-\frac{1}{D t} \bigl(p'-p\bigr)^2
               \right] W_{\mathrm{red}}(0,x,p')
\\
& = \left(\frac{1}{4 \pi D t}\right)^{\frac{1}{2}}
        \int dp' \exp\left[-\frac{1}{4 D t} 
                    (p'-p)^2 
                    \right] W_{\mathrm{red}}(0,x,p')
\end{align*}
If this result is re-expressed in terms
of the density matrix one recovers Eq.~(\ref{eq:  quasiXdiag})
(with $\Lambda = D/\hbar^2$). It follows that the discussion in
the last section only applies to the situation when
$t \ll
\tcrit$, before there has been any significant degree of
smoothing in the
$x$-direction. The question we now have to consider is whether
there is a tendency for the distribution of Bohmian velocities
to become generically quasi-classical when $t > \tcrit$.}

The result established by Halliwell and Zoupas~\cite{HalliB}
provides some preliminary indication that such an outcome
might be expected.  Halliwell and Zoupas show that, in the
case of the Caldeira-Leggett model as applied to a free
particle, with negligible dissipation, the Wigner function
becomes strictly non-negative once $t \ge
\left(3/16\right)^{1/4}\tloc$, where
$\tloc$ is the localization time given by
\begin{equation}
  \tloc = \left(\frac{\hbar}{\gamma \bolt T}\right)^{\frac{1}{2}}
\label{eq:  tlocDef}
\end{equation}
It is easily seen that this is also true for the case of
a harmonically bound particle considered here (provided 
$\omega \tloc, \gamma \tloc \ll 1$, so that the
approximation of Eq.~(\ref{eq:  CLWigProx}) is still valid
at $t=\tloc$).  More generally, what is essentially the same
argument shows that, whenever the Wigner function propagator
takes the form specified by Eq.~(\ref{eq:  RedWigProp}), $W_{\mathrm{red}}$
becomes strictly non-negative once
\begin{equation*}
  \det \mathbf{M}(t) \ge \hbar^2
\end{equation*}
It follows from Eqs.~(\ref{eq:  bveTermsRhoRed}) 
 and~(\ref{eq:  RedWigDefA}) 
that
\begin{equation}
  \bve(t,x) = 
  \frac{\int dp \, p W_{\mathrm{red}}(t,x,p)}{m\int dp\,
  W_{\mathrm{red}}(t,x,p)}
\label{eq:  bveTermsWig}
\end{equation}
If $W_{\mathrm{red}}$ is non-negative, and if it is negligible outside
the region enclosed  by the classical energy surface $p=\pm p_{\mathrm{cl}}(x)$,
then it can be seen from this expression that $\bve$ will lie within, or close
to the classically permitted range  for the mean velocity:  $-p_{\mathrm{cl}}(x)/m
\lesssim \bve(t,x) \lesssim p_{\mathrm{cl}}(x)/m$.

We now illustrate this phenomenon by calculating $W_{\mathrm{red}}$ when
$\ket{\psi_{\mathrm{sys}}}$ is an
approximate energy eigenstate of the form specified by 
Eq.~(\ref{eq:  ProxESteDef}).  It turns out that for such states $\bve$
typically comes to lie within the classically permitted range of values at times
significantly earlier than $\tloc$.

It was shown in ref.~\cite{self1}, using the WKB approximation, that
\begin{equation*}
  \overlap{x}{\psi_{\mathrm{sys}}} 
\approx
  i \left( \exp\left[- \frac{i}{\hbar} S(x)\right] g_{-}(x)
           - \exp \left[\frac{i}{\hbar} S(x) \right] g_{+} (x)
\right) 
\end{equation*} 
provided that $x$ is not close to one of the classical turning
points at
$x=\pm \amp$.  In this expression
\begin{equation*}
  S(x)  = \int_{-\amp}^{x} dx' \, p_{\mathrm{cl}} (x')+\frac{h}{8}
\end{equation*} 
and
\begin{align*}
& g_{\pm} (x)  \\& =  
\begin{cases} \left(\frac{\omega}{2 \pi
p_{\mathrm{cl}}(x)}\right)^{\frac{1}{2}}
                  \sum_{r=-\frac{\Delta n}{2}}^{\frac{\Delta
n}{2}} c_r \exp\left[\pm i r \sin^{-1}
\Bigl(\frac{x}{\amp}\Bigr)\right]
\qquad &\text{if\ }-\amp<x<\amp \\ 0
\qquad &\text{otherwise}
               \end{cases}
\end{align*}
This gives, for the reduced Wigner function at $t=0$,
{\allowdisplaybreaks
\begin{align*} 
& W_{\mathrm{red}} (0,x,p) \notag
\\ & \approx \frac{1}{h} 
   \int dy \, \exp\left(\frac{i}{\hbar} p y\right) \notag
\\ & \hspace{0.5 in} \times
          \left\{
\exp\left[-\frac{i}{\hbar}S\left(x-\frac{y}{2}\right)\right]
                      g_{-}
\left(x-\frac{y}{2}\right)                  
-\exp\left[\frac{i}{\hbar}S\left(x-\frac{y}{2}\right)\right]
                      g_{+} \left(x-\frac{y}{2}\right)
         \right\} \notag
\\ & \hspace{0.5 in} \times
          \left\{
\exp\left[\frac{i}{\hbar}S\left(x+\frac{y}{2}\right)\right]
                      g_{-}^{*} \left(x+\frac{y}{2}\right)
-\exp\left[-\frac{i}{\hbar}S\left(x+\frac{y}{2}\right)\right]
                      g_{+}^{*} \left(x+\frac{y}{2}\right)
         \right\}
\end{align*} 
Substituting this expression into  
Eq.~(\ref{eq:  RedWigProp}) and  carrying out the 
$p'$-integration
gives
\begin{align} 
& W_{\mathrm{red}}(t,x,p) \notag
\\  &  \approx \frac{\sqrt{\Delta}}{h 
\sqrt{\pi b}}
  \int dx' dy' \, \notag
\\  & \hspace{0.15 in} \times
    \exp \left[ -\frac{1}{4 \hbar^2 b} {y'}^2-\frac{\Delta}{b}
(x'-x)^2 
               + \frac{i}{\hbar} \left( p-\frac{c}{b}
(x'-x)\right) {y'}\right] \notag
\\ & \hspace{0.15 in} \times
          \left\{
\exp\left[-\frac{i}{\hbar}S\left(x'-\frac{{y'}}{2}\right)\right]
                      g_{-} \left(x'-\frac{{y'}}{2}\right)
-\exp\left[\frac{i}{\hbar}S\left(x'-\frac{{y'}}{2}\right)\right]
                      g_{+} \left(x'-\frac{{y'}}{2}\right)
         \right\} \notag
\\ & \hspace{0.15 in} \times
          \left\{
\exp\left[\frac{i}{\hbar}S\left(x'+\frac{{y'}}{2}\right)\right]
                      g_{-}^{*}
\left(x'+\frac{{y'}}{2}\right)           
-\exp\left[-\frac{i}{\hbar}S\left(x'+\frac{{y'}}{2}\right)\right]
                      g_{+}^{*} \left(x'+\frac{{y'}}{2}\right)
         \right\}
\label{eq:  RedWigCalcA}
\end{align}
where we have assumed that $t$ is sufficiently small to justify
the approximation $\mathbf{A}^{-1}(t) \approx \boldsymbol{1}$, and 
where we have set}
\begin{equation}
 \mathbf{M}^{-1}  = \begin{pmatrix} a & c\\ c & b\end{pmatrix}
\hspace{1 in}
 \det \mathbf{M}^{-1}  = \Delta  
\label{eq:  MInvforCL}
\end{equation}
In order to evaluate this expression we note, first of all,
that the functions $g_{\pm}$ are slowly-varying~\cite{self1}. We
may therefore write
\begin{equation*}
 g_{+}\left(x'\pm\frac{y'}{2}\right) \approx g_{+}(x)
 \hspace{0.5 in}
 g_{-}\left(x'\pm\frac{y'}{2}\right) \approx g_{-}(x)
\end{equation*}
provided that the
Gaussian peaks are sufficiently narrow, and provided that 
$x$ is not too close to one of the classical 
turning points. It will also be convenient  to write these functions 
in modulus-argument form:
\begin{equation*}
 g_{\pm}   = \sqrt{\rho_{\pm}}
e^{i
\phi_{\pm}}
\end{equation*}
Finally we make the approximation
\begin{equation}
 \frac{1}{\hbar} S\left(x'\pm \frac{y'}{2}\right)
 \approx
 \frac{1}{\hbar} S(x)
 +\frac{1}{\hbar} p_{\mathrm{cl}}(x)
 \left( x'\pm\frac{y'}{2} - x\right) 
 +\frac{1}{2\hbar} {p'}_{\mathrm{cl}}(x)
 \left( x'\pm\frac{y'}{2} - x\right)^2
\label{eq:  SProx}
\end{equation}
This approximation will be
justified provided
\begin{equation}
\frac{1}{\hbar} \left|{p''}_{\mathrm{cl}}(x)\right|
\left(\frac{b}{\Delta}\right)^{\frac{3}{2}} \ll 1 \hspace{0.3 in}
\text{and} \hspace{0.3 in}
\frac{1}{\hbar} \left|{p''}_{\mathrm{cl}}(x)\right|
\left(4 \hbar^2 b\right)^{\frac{3}{2}} \ll 1
\label{eq:  ValidCondA}
\end{equation}
or, using the fact that
$\left|{p''}_{\mathrm{cl}}(x)\right| \sim m
/(\omega \amp)$ everywhere except in the vicinity of the
classical turning points,
\begin{equation}
\frac{m \omega b^{\frac{3}{2}}}{\hbar  \Delta^{\frac{3}{2}}}
\ll \amp \hspace{0.3 in}
\text{and} \hspace{0.3 in}
8 m \omega \hbar^2 b^{\frac{3}{2}}  \ll \amp
\label{eq:  SProxCond}
\end{equation}
Making these approximations in Eq.~(\ref{eq:  RedWigCalcA}) and
carrying out the Gaussian integrations gives, after a certain
amount of algebra,
\begin{equation}
  W_{\mathrm{red}}(t,x,p) \approx W_{\mathrm{cl}}
(t,x,p)+W_{\mathrm{osc}}(t,x,p)
\label{eq:  WigDecomp}
\end{equation}
where
\begin{multline*}
  W_{\mathrm{cl}} (t,x,p)
= \frac{\sigma_{-}}{\sqrt{\pi} }
  \exp\left[ - \sigma_{-}^{2}
          \bigl( p + p_{\mathrm{cl}}(x)\bigr)^2
          \right] \rho_{-} (x)\\ 
   + \frac{\sigma_{+}}{ \sqrt{\pi }}
  \exp\left[ - \sigma_{+}^{2}
          \bigl( p - p_{\mathrm{cl}}(x)\bigr)^2
          \right] \rho_{+} (x)
\end{multline*}
and
\begin{multline}
W_{\mathrm{osc}}(t,x,p) =
\left(\frac{4 \hbar \sigma_1 \sigma_2 \sqrt{\Delta}}{\pi } 
\right)^{\frac{1}{2}}
 \exp\left[-\sigma_{2}^{2} \bigl(p+\beta 
            p_{\mathrm{cl}}(x)\bigr)^2
            -\sigma_{1}^{2} 
            \bigl(p_{\mathrm{cl}}(x)\bigr)^2\right]
\\ \times
\cos \left[ \frac{2}{\hbar} S(x) + \chi(t,x,p)\right]
\bigl(\rho_{-}(x) \rho_{+}(x)
\bigr)^\frac{1}{2}
\label{eq:  WigOsc}
\end{multline}
and where we have set
\begin{align*}
\rho_{\pm} & = \left|g_{\pm}\right|^2 \\
\sigma_{\pm}^{2} & = \frac{ \Delta}{ a \pm 2 c p'_{\mathrm{cl}}(x)
   +b \bigl(p'_{\mathrm{cl}}(x)\bigr)^2 } \\
\sigma_{1}^{2} & =
\frac{\Delta}{ \hbar^2 a \Delta+b
\bigl(p'_{\mathrm{cl}}(x)\bigr)^2} 
\\
\sigma_{2}^{2} & =
\frac{\hbar^2 a \Delta + b 
   \bigl(p'_{\mathrm{cl}}(x)\bigr)^2}{
   \hbar^2  a^2 +\left(1-2
     \hbar^2 c^2 + \hbar^4 \Delta^2\right)
     \bigl(p'_{\mathrm{cl}}(x)\bigr)^2
     +\hbar^2  b^2 \bigl(p'_{\mathrm{cl}}(x)\bigr)^4 }
\\
\beta & =
\frac{
        c\left(1+\hbar^2\Delta\right)
        p'_{\mathrm{cl}}(x) 
        }{
        \hbar^2 a \Delta + b 
        \bigl(p'_{\mathrm{cl}}(x)\bigr)^2
        }
\end{align*}
$\chi(t,x,p)$ is a phase whose functional form is unimportant
for present purposes.

$W_{\mathrm{cl}}$ is non-negative, and it is concentrated 
on the classical energy surface at $p=\pm p_{\mathrm{cl}}$.  
It is therefore a possible classical phase space
probability distribution describing a particle of energy
$E_{\bar{n}}$, with $\rho_{+}$ (respectively $\rho_{-}$)
being the probability density function for the particle
to be located at $x$ and moving to the right (respectively
left).
On the other hand the term $S/\hbar$ in the argument of the 
cosine means that $W_{\mathrm{osc}}$ is very rapidly oscillating.
$W_{\mathrm{osc}}$ is the term responsible for the tendency of the
Wigner function to swing negative.  It may
therefore be regarded as the
quantum mechanical correction to the classical distribution.

It can be seen from Eq.~(\ref{eq:  WigOsc}) that $W_{\mathrm{osc}}$ will 
become
negligible  once
\begin{equation*}
  \sigma_{1} p_{\mathrm{cl}}(x) \gg 1
\end{equation*}
In that case $W_{\mathrm{red}}\approx W_{\mathrm{cl}}$ and, in view of 
Eq.~(\ref{eq:  bveTermsWig}),
\begin{equation*}
 \bve (x) \approx \frac{p_{\mathrm{cl}}(x)}{m} 
 \left( \frac{\rho_{+}(x) - \rho_{-}(x)}{\rho_{+}(x)+\rho_{-}(x)}\right)
\end{equation*}
from which it follows that $\bve$ lies within the classical range 
\begin{equation*}
  - \frac{p_{\mathrm{cl}}(x)}{m}
\le \bve (x)
\le
  \frac{p_{\mathrm{cl}}(x)}{m}
\end{equation*}

Specialising to the case of the Caldeira-Leggett model it can be seen from
Eqs.~(\ref{eq:  MforCL}), (\ref{eq:  MInvforCL}) 
and~(\ref{eq:  SProxCond}) that the condition for the approximation
of Eq.~(\ref{eq:  SProx})
to  be valid is
\begin{equation}
  \left(\omega \tloc\right)^{\frac{4}{3}}
\ll \frac{t}{\tcrit}
\ll \left(\frac{\amp}{\deBrog}\right)^{\frac{4}{9}}
\label{eq:  ValidCondB}
\end{equation}
where $\tcrit$ is the time at which the
smearing in the
$x$-direction becomes significant [see Eq.~(\ref{eq:  tcDef})], 
$\tloc$ is the localisation time [see
Eq.~(\ref{eq:  tlocDef})], and  $\deBrog=\hbar / (m 
\omega \amp)$, as before.
We also have
\begin{equation*}
 \sigma_{1}^{2} p_{\mathrm{cl}}^2
=  \frac{D  t^3 p_{\mathrm{cl}}^2}{3 m^2 \hbar^2}
\left(1+ \frac{4 D^2 t^6 {p'}_{\mathrm{cl}}^2 }{9 m^4 \hbar^2}\right)^{-1}
\end{equation*}
Provided that $x$ is not too close to one of the classical turning points
at 
$x=\pm \amp$ we have $p_{\mathrm{cl}}(x)
\sim m
\omega \amp$ and
${p'}_{\mathrm{cl}}(x) \sim m \omega$.  Consequently
\begin{equation*}
 \sigma_{1}^{2} p_{\mathrm{cl}}^2
\sim \frac{\left(\frac{t}{\tcrit}\right)^3}{
1+ 4 \left(\frac{\deBrog }{\amp}\right)^2
   \left( \frac{t}{\tcrit}\right)^6}
\end{equation*}
The fact that $\ket{\psi_{\mathrm{sys}}}$ is highly excited means
that $\amp \gg \deBrog$.  Taking into account  inequalities~(\ref{eq:  ValidCondB})
we conclude that, in the case of the Caldeira-Leggett model,
$W_{\mathrm{osc}}$ is  negligible, and $\bve$ is approximately within the classical
range of values, once
$t \gg
\tcrit$.

Finally, we remark that it follows from
Eqs.~(\ref{eq:  tcDef})
and~(\ref{eq:  tlocDef}) that
\begin{equation*}
  \frac{\tcrit}{\tloc} = \left(\frac{3\hbar \gamma k T}{8
E_{\bar{n}}^{2}}\right)^{\frac{1}{6}}
\end{equation*}
where $E_{\bar{n}}=(\bar{n}+1/2) \hbar \omega = (1/2)m \omega^2\amp^2$
is the mean energy.  We see from this that, in the case of a macroscopic body, 
$\tcrit$
is typically
$\ll
\tloc$. As we mentioned above, $\tloc$ is the time at which 
the Wigner function becomes strictly non-negative,
for every possible choice of initial state~\cite{HalliB}.
However, the argument just given shows that, for states of the form
specified by Eq.~(\ref{eq:  ProxESteDef}), the Wigner function 
typically becomes approximately non-negative very much sooner
than this, and approximate non-negativity is enough to 
ensure that $\bve$ lies approximately within the classical range
of values.

\section{Calculation of $\vb^{(\alpha)}(t,x,x_1,\dots,x_N)$}
\label{sec:  QuantBrown}
The requirement that $|\bve| \le p_{\mathrm{cl}}/m$ is a necessary
condition for the Bohmian trajectories to be quasi-classical.  However,
it is clearly not sufficient.
We therefore  need to turn from
the ensemble-averaged quantity $\bve(t,x)$ to the Bohmian velocity
itself, $\vb^{(\alpha)}(t,x,x_1,\dots,x_N)$. 

In order to calculate $\vb^{(\alpha)}$ it is necessary to resolve the
ambiguity mentioned in the Introduction, arising from the fact that the 
density matrix $\hat{\rho}_{\mathrm{bath}}$ [see Eq.~(\ref{eq:  RhoBatDef})],
describing the initial state of the heat bath, does not uniquely 
determine a corresponding ensemble [see the discussion in the paragraph following
Eq.~(\ref{eq:  BathEnsemble})].
One obvious choice is to represent $\hat{\rho}_{\mathrm{bath}}$ in terms of
eigenstates of the heat bath Hamiltonian:
\begin{equation}
  \hat{\rho}_{\mathrm{bath}} =  \sum_{E} 
  p_{E} \ket{E}\bra{E}
\label{eq:  rhoEnvA}
\end{equation}
where $\ket{E}$ is the eigenstate of $\hat{H}_{\mathrm{bath}}$ with
eigenvalue
$E$ and  $p_{E} = \mathscr{N} \exp\left[-E/(\bolt T)\right] $
[\emph{c.f.}\ Eq.~(\ref{eq:  RhoBatDef})].
However, we will find it more convenient to use the coherent
state representation
\begin{multline}
  \hat{\rho}_{\mathrm{bath}}
= \int d\bar{x}_1 d\bar{p}_1 \dots d\bar{x}_N d\bar{p}_N \,
   P\left(\bar{x}_1,\bar{p}_1,\dots,\bar{x}_N,\bar{p}_N\right)
\\ \times
   \ket{\bar{x}_1,\bar{p}_1,\dots,\bar{x}_N,\bar{p}_N}
   \bra{\bar{x}_1,\bar{p}_1,\dots,\bar{x}_N,\bar{p}_N}
\label{eq:  rhoEnvB}
\end{multline}
where $P$ is the thermal Glauber-Sudarshan 
$P$-function~\cite{Hil,Lee,Leon,Glauber,Sudar}
\begin{equation*}
P\left(\bar{x}_1,\bar{p}_1,\dots,\bar{x}_N,\bar{p}_N\right) 
= \prod_{r=1}^{N} \left(
\frac{\left(e^{\beta_r}-1\right)}{h}
\exp\left[-\frac{1}{2}\left(e^{\beta_r}-1\right)
 \left(\frac{1}{\lambda_{r}^{2}} \bar{x}_r^2
+\frac{\lambda_{r}^{2}}{\hbar^2} \bar{p}_r^2\right)\right]
\right)
\end{equation*}
and 
$\ket{\bar{x}_1,\bar{p}_1,\dots,\bar{x}_N,\bar{p}_N}$ is the  coherent state with
$x$-space wave function
\begin{multline*}
  \overlap{x_1,\dots,x_N}{\bar{x}_1,\bar{p}_1,\dots,\bar{x}_N,\bar{p}_N} \\
= \prod_{r=1}^{N}
    \left( 
      \left(\frac{m_{r} \omega_{r}}{\pi \hbar}
      \right)^{\frac{1}{4}}
      \exp 
      \left[  -\frac{m_{r} \omega_{r}}{2 \hbar} 
           \left(     x_r-\bar{x}_r
           \right)^2
            +   \frac{i}{\hbar} \bar{p}_r x_r 
            - \frac{i}{ 2\hbar}\bar{p}_{r}\bar{x}_{r}
       \right]
      \right)
\end{multline*}
$\beta_r$ denotes the ratio
$\hbar \omega_r / (\bolt T)$, .

Eqs.~(\ref{eq:  rhoEnvA}) and~(\ref{eq:  rhoEnvB}) are completely
equivalent ways of writing the density matrix.  Consequently, from the
point of view of the Copenhagen interpretation it makes no difference,
whether we take the ensemble corresponding to $\hat{\rho}_{\mathrm{bath}}$
to consist of the states $\ket{E}$ with (discrete) probability
distribution
$p_{E}$, or whether we take it to consist of the 
states $\ket{\bar{x}_1,\bar{p}_1,\dots,\bar{x}_N,\bar{p}_N}$ with
(continuous) probability distribution
$P(\bar{x}_1,\bar{p}_1,\dots,\bar{x}_N,\bar{p}_N)$. However, from the point
of view of the Bohm interpretation it does make a  difference;
since in the former case, but not in the latter, the Bohmian velocity of the
$r^{\mathrm{th}}$ heat bath particle is certainly zero at $t=0$.
In the following we will assume that the  ensemble giving rise to
$\hat{\rho}_{\mathrm{bath}}$ consists
of coherent states, since the integrals are then much easier to 
calculate.  The question, as to whether choosing a different ensemble
would require our conclusion to be modified, we defer to a future 
investigation.

 Eq.~(\ref{eq:  SysHtBth}), giving the state of system$+$heat bath
at time $t$, thus becomes
\begin{equation*}
  \ket{\Psi_{\alpha}(t)} = e^{- i t \hat{H}/\hbar} \bigl(
  \ket{\psi_{\mathrm{sys}}} \otimes 
  \ket{\bar{x}_{1},\bar{p}_1,\dots,\bar{x}_N,\bar{p}_{N}}\bigr)
\end{equation*}
where the label $\alpha$  now denotes the
$2N$-tuple
$(\bar{x}_1,
\dots,
\bar{p}_N)$.
Let $W_{\alpha}$ be the corresponding Wigner function:
\begin{align}
& W_{\alpha}(t,x,p,x_{1},p_{1},\dots,x_{N},p_{N}) \notag \\
& = \frac{1}{h^N}\int dy dy_1 \dots dy_N \,
     \exp\biggl[\frac{i}{\hbar}\biggl(p y+ \sum_{r=1}^{N} p_{r} y_{r}\biggr)
\biggr] \notag \\
& \hspace{0.2 in}  \times 
   \overlap{x-\tfrac{y}{2},x_1-\tfrac{y_1}{2},\dots,x_N-\tfrac{y_N}{2}}{
     \Psi_{\alpha}(t)}
   \overlap{
     \Psi_{\alpha}(t)}{x+\tfrac{y}{2},x_1+\tfrac{y_1}{2},\dots,x_N+\tfrac{y_N}{2}}
\label{eq:  WPsiDef}
\end{align}
and define
\begin{equation}
\widetilde{W}_{\alpha}(t,x,p,x_1,\dots,x_N)
= \int dp_1 \dots dp_N \, W_{\alpha}(t,x,p,x_1,p_1,\dots,x_N,p_N)
\label{eq:  Wtilde}
\end{equation}
Eqs.~(\ref{eq:  BVelDef}), (\ref{eq:  WPsiDef}) 
and~(\ref{eq:  Wtilde}) imply
\begin{equation}
  \vb^{(\alpha)}(t,x,x_1,\dots,x_N)
= \frac{\int dp  \, p \, \widetilde{W}_{\alpha}(t,x,p,x_1,\dots,x_N)
      }{m \int dp \, \widetilde{W}_{\alpha}(t,x,p,x_1,\dots,x_N)}
\label{eq:  vbTermsWigB}
\end{equation}
This equation is of a similar form to Eq.~(\ref{eq:  bveTermsWig}).  It should,
however, be noted that $\vb^{(\alpha)}$ is the \emph{actual} Bohmian velocity, whereas
the quantity
$\bve$ given by  Eq.~(\ref{eq:  bveTermsWig}) is  only an average.

In order to calculate this quantity we follow
Halliwell and Yu~\cite{HalliC}, and use  the fact that, because the
Hamiltonian is quadratic in the positions and momenta, the Wigner function
propagates in  the same way as a 
classical phase space distribution.

It is convenient to employ the vector notation:
\begin{equation*}
  \boldsymbol{\eta} = \begin{pmatrix} x \\ p\end{pmatrix}
  \hspace{0.5 in}
  \boldsymbol{\eta}_r = \begin{pmatrix} x_r \\ p_r \end{pmatrix}
\end{equation*}
Let $\boldsymbol{\eta}(t)$, $\boldsymbol{\eta}_{r}(t)$ be solutions 
to the classical equations of motion, which result from the classical analogue
of the Hamiltonian of Eq.~(\ref{eq:  BrownHamDef}).
Since the Hamiltonian is quadratic we have
\begin{align}
  \boldsymbol{\eta}(t) & = \mathbf{A}(t) \boldsymbol{\eta}(0) + 
    \sum_{r=1}^{N} \mathbf{B}_r (t) \boldsymbol{\eta}_{r}(0) 
\label{eq:  etaEvol} \\
  \boldsymbol{\eta}_r(t) & = \mathbf{C}_r (t) 
  \boldsymbol{\eta}(0) + \sum_{r'=1}^N \mathbf{D}_{r r'} (t) 
  \boldsymbol{\eta}_{r'}(0) 
\label{eq:  etarEvol} 
\end{align}
for suitable matrices $\mathbf{A}(t)$, $\mathbf{B}_r (t)$, $\mathbf{C}_r (t)$,
$\mathbf{D}_{r r'} (t)$.
In Appendix~\ref{app:  ExactQBrown} we give exact, closed form expressions
for these matrices. In Appendix~\ref{app:  RedWigProp} we show how the matrix
$\mathbf{M}$ appearing in Eq.~(\ref{eq:  RedWigProp}) can be expressed in terms
of them.

We can use these matrices to propagate $W_{\alpha}$ forward in time~\cite{Hil,Lee}:
\begin{multline}
  W_{\alpha}(t,\boldsymbol{\eta},\boldsymbol{\eta}_1, \dots,
\boldsymbol{\eta}_N)\\ = W_{\alpha}\biggl(0,\mathbf{A}(-t) \boldsymbol{\eta}
+\sum_{r}
\mathbf{B}_{r}(-t)\boldsymbol{\eta}_{r},
\mathbf{C}_{1}(-t)\boldsymbol{\eta} +\sum_{r} \mathbf{D}_{1 r}(-t)
\boldsymbol{\eta}_{r},\\ \dots,
\mathbf{C}_{N}(-t)\boldsymbol{\eta} +\sum_{r} \mathbf{D}_{N r}(-t)
\boldsymbol{\eta}_{r}
\biggr)
\label{eq:  WEvolA}
\end{multline}
Also
\begin{equation}
  W_{\alpha}(0,\boldsymbol{\eta},\boldsymbol{\eta}_1,\dots,\boldsymbol{\eta}_N)
= W_{\mathrm{sys}}(\boldsymbol{\eta})
W_{\mathrm{bath}}^{(\alpha)}(\boldsymbol{\eta}_1,\dots,\boldsymbol{\eta}_N)
\label{eq:  WTimeZero}
\end{equation}
where $W_{\mathrm{sys}}$, $W_{\mathrm{bath}}^{(\alpha)}$ are the Wigner functions
corresponding to 
$\ket{\psi_{\mathrm{sys}}}$, 
$\ket{\bar{\boldsymbol{\eta}}_1,\dots, \bar{\boldsymbol{\eta}}_N}$ respectively.
We have~\cite{Hil,Lee,Leon}
\begin{equation}
W_{\mathrm{bath}}^{(\alpha)}(\boldsymbol{\eta}_1,\dots,\boldsymbol{\eta}_N)
=\frac{2^N}{h^N} 
  \exp \left[  \sum_{r=1}^{N} 
              (\boldsymbol{\eta}_r - \bar{\boldsymbol{\eta}}_r)^{\mathrm{T}}
              \boldsymbol{\Lambda}_r
              (\boldsymbol{\eta}_r - \bar{\boldsymbol{\eta}}_r)
      \right]
\label{eq:  Wbath}
\end{equation}
where
\begin{equation}
  \boldsymbol{\Lambda}_r = 
  \frac{1}{\hbar}\begin{pmatrix} m_{r} \omega_{r} & 0 \\
                   0 & \frac{1}{m_{r} \omega_{r}}
  \end{pmatrix}
\label{eq:  LambdaDef}
\end{equation}
Using Eqs.~ (\ref{eq:  Wtilde}), (\ref{eq:  WEvolA}),
(\ref{eq:  WTimeZero}), (\ref{eq:  Wbath}) we deduce
\begin{equation}
  \widetilde{W}_{\alpha}(t,\boldsymbol{\eta},x_1,\dots,x_N)
= \int d^{2}\boldsymbol{\eta}' \,
\tilde{G}
(t,\boldsymbol{\eta},x_1,\dots,x_N|\boldsymbol{\eta}')
W_{\mathrm{sys}}(\boldsymbol{\eta}')
\label{eq:  WtildeTermsWsys}
\end{equation}
where
{\allowdisplaybreaks
\begin{align}
&  \tilde{G} (t,\boldsymbol{\eta},x_1,\dots,x_N|\boldsymbol{\eta}')
\notag \\
& 
= \frac{2^{N-2}}{\pi^2 h^N}
  \int d^{2}\boldsymbol{\xi} dp_1 \dots dp_N \,
   \exp\biggl[i\boldsymbol{\xi}^{\mathrm{T}}
             \biggl(\boldsymbol{\eta}' -\mathbf{A}(-t) \boldsymbol{\eta}
                   -\sum_{r=1}^{N} \mathbf{B}_r(-t) \boldsymbol{\eta}_r
             \biggr)
       \biggr.
\notag \\
& \hspace{0.18 in}
       \biggl. -
             \sum_{r,r',r''=1}^{N}
             \Bigl( \mathbf{C}_{r}(-t)\boldsymbol{\eta}+\mathbf{D}_{r
                    r'}(-t) 
                    \boldsymbol{\eta}_{r'}-\boldsymbol{\bar{\eta}}_r
             \Bigr)^{\mathrm{T}}
             \boldsymbol{\Lambda}_r
             \Bigl( \mathbf{C}_{r}(-t)\boldsymbol{\eta}+\mathbf{D}_{r
                     r''}(-t) 
                    \boldsymbol{\eta}_{r''}-\boldsymbol{\bar{\eta}}_r
             \Bigr)
       \biggr]
\label{eq:  GIntA}
\end{align}
Carrying out the Gaussian integrations in Eq.~(\ref{eq:  GIntA}), and using
the results in Appendix~\ref{app:  ExactQBrown}, gives an exact, closed form
expression for $\tilde{G}$.  However, the expression is
rather complicated, due to the fact that $\mathbf{D}_{r r'}$ couples
together the different oscillators constituting the heat bath.  We will
therefore confine ourselves to the case when the interaction betweeen heat
bath and system is weak.  It is shown in Appendix~\ref{app:  ExactQBrown}
that we may then  approximate
\begin{align}
  \mathbf{A}(t) & \approx \begin{pmatrix} \cos \omega t & 
                                    \frac{1}{m \omega} \sin\omega t \\
                                    - m \omega \sin \omega t &
                                    \cos \omega t
                    \end{pmatrix}
\label{eq:  AProx}\\
  \mathbf{B}_{r}(t) & \approx \frac{ \kappa_r}{m m_r \omega \omega_r}
                   \begin{pmatrix} m_r \dot{h}_{r}^{(0)}(t)  & 
                   h_{r}^{(0)}(t)
                           \\ 
                         m m_r \ddot{h}_{r}^{(0)}(t) & 
                         m\dot{h}_{r}^{(0)}(t)
           \end{pmatrix}
\label{eq:  BProx}\\
  \mathbf{C}_{r}(t) & \approx \frac{\kappa_r}{m m_r \omega \omega_r}
                   \begin{pmatrix} m \dot{h}_{r}^{(0)}(t)  & 
                   h_{r}^{(0)} (t)
\\
                         m m_r \ddot{h}_{r}^{(0)}(t) & 
                         m_r\dot{h}_{r}^{(0)}(t)
           \end{pmatrix} 
\label{eq:  CProx}
\\
  \mathbf{D}_{r r'}(t) & \approx 
             \delta_{r r'} \mathbf{D}_{r}(t)
\label{eq:  DProx}
\\
\intertext{where}
  h_{r}^{(0)} (t) & = -\frac{\omega_{r} \sin \omega t  -\omega 
     \sin \omega_{r} t}{\omega_{r}^{2} - \omega^{2}}
\notag
\\
\intertext{and}
  \mathbf{D}_{r}(t) & =
  \begin{pmatrix} \cos \omega_r t & \frac{1}{m_r \omega_r} \sin \omega_r t
            \\
                  - m_r \omega_r \sin \omega_r t & \cos \omega_r t
  \end{pmatrix}
\notag
\end{align}
If we also assume that $\omega t \ll 1$, then we can further approximate
\begin{equation*}
\mathbf{A}(t) \approx \mathbf{1}  \hspace{0.5 in}
\text{and} \hspace{0.5 in}
  h_{r}^{(0)} (t)  \approx - \frac{\omega \left(  \omega_{r} t- 
                                   \sin \omega_r t\right)}{\omega_{r}^{2}}
\end{equation*}
Using these results in 
Eq.~(\ref{eq:  GIntA}) and carrying out the integrations gives
\begin{align}
  \tilde{G}
(t,\boldsymbol{\eta},x_1,\dots,x_N|\boldsymbol{\eta}') & \approx
  \mathrm{const.} \times
  \exp\biggl[-\sum_{r=1}^{N} \frac{m_{r} \omega_{r}}{\hbar}
          \bigl(x_r - q_r(\boldsymbol{\eta})\bigr)^2
      \biggr]
\notag
\\ & \hspace{0.5 in}\times \exp\biggl[
             -\left(\boldsymbol{\eta}'-\boldsymbol{\eta}-
                    \boldsymbol{\delta}
              \right)^{\mathrm{T}}
              \widetilde{\mathbf{M}}^{-1}(t)
              \left(\boldsymbol{\eta}'-\boldsymbol{\eta}-
                    \boldsymbol{\delta}
              \right)
      \biggr]
\label{eq:  GExpress}
\end{align}
where
{\allowdisplaybreaks
\begin{align}
q_r(\boldsymbol{\eta}) & =
\Bigl(\mathbf{D}_{r}(t) \bigl(\boldsymbol{\bar{\eta}}_r
                    - \mathbf{C}_{r}(-t)
                            \boldsymbol{\eta}\bigr)
                      \Bigr)_{1}
\label{eq:  qrDef}
\\
\boldsymbol{\delta} & =
  \sum_{r=1}^{N} \left( \mathbf{B}_r (-t)
  \begin{pmatrix} x_r \\
            \Bigl( \mathbf{D}_{r}(t)
                  \bigl(\bar{\boldsymbol{\eta}}_r
                     - \mathbf{C}_r(-t) \boldsymbol{\eta}
                  \bigr)
            \Bigr)_2
  \end{pmatrix} \right)
\label{eq:  BoldDeltaDef}
\\
\widetilde{\mathbf{M}}(t)
& = 2\hbar \int_{0}^{\infty} d \omega' \,
       I(\omega')
       \begin{pmatrix} \frac{\left(\omega' t-\sin \omega' t\right)^2}{
                      m^2 {\omega'}^4} &
                     - \frac{\left(\omega' t-\sin \omega' t\right)
                           \left(1-\cos\omega' t\right)}{m {\omega'}^3}
                \\
                      -\frac{\left(\omega' t-\sin \omega' t\right)
                           \left(1-\cos\omega' t\right)}{m {\omega'}^3} &
                        \frac{\left(1-\cos\omega' t\right)^2}{
                           {\omega'}^2}
       \end{pmatrix}
\label{eq:  MTwiddleDef}
\end{align}
and $I(\omega')$ is the spectral density function defined by 
Eq.~(\ref{eq:  IomegDef}).
Eqs.~(\ref{eq:  WtildeTermsWsys}) and~(\ref{eq:  GExpress}) then imply}
\begin{multline*}
\widetilde{W}_{\alpha}\left(t,\boldsymbol{\eta},x_1,\dots,x_N\right)
\approx
  \left(\frac{m_{1}\omega_{1} \dots m_{N}\omega_{N} }{\pi^{N+2} \hbar^N
  \det \widetilde{\mathbf{M}}(t)}
  \right)^{\frac{1}{2}}
\exp\biggl[-\sum_{r=1}^{N} \frac{m_{r} \omega_{r}}{\hbar} 
\bigl(x_r-q_r(\boldsymbol{\eta})\bigr)^2\biggr]
\\ \times
\int d^{2}\boldsymbol{\eta}' \, 
     \exp \left[-(\boldsymbol{\eta}'-\boldsymbol{\eta})^{\mathrm{T}}
                  \widetilde{\mathbf{M}}^{-1}(t) 
                  (\boldsymbol{\eta}'-\boldsymbol{\eta})
           \right] W_{\mathrm{sys}}(\boldsymbol{\eta}')
\end{multline*}
where we have set $\boldsymbol{\delta}\approx\mathbf{0}$, which 
will be justified if the coupling between system and heat bath is
sufficiently weak.
Define
\begin{equation}
  \widetilde{\mathbf{M}}^{-1} 
   = \begin{pmatrix} \tilde{a} & \tilde{c} \\ 
                     \tilde{c} & \tilde{b}
      \end{pmatrix}
\hspace{0.5 in} \text{and}  \hspace{0.5 in}
  \det\widetilde{\mathbf{M}}^{-1}  = \tilde{\Delta}
\label{eq:  DeltaTwiddleDef}
\end{equation}
We then have, by essentially the same argument as the one leading to 
Eq.~(\ref{eq:  WigDecomp}), that if
\begin{equation}
  \frac{m \omega {\tilde{b}}^{\frac{3}{2}}}{\hbar
{\tilde{\Delta}}^{\frac{3}{2}}}
  \ll \amp \hspace{0.5 in} \text{and} \hspace{0.5 in}
  8 m \omega \hbar^2 {\tilde{b}}^{\frac{3}{2}} \ll \amp
\label{eq:  WigTwidProxCond}
\end{equation}
then
\begin{equation}
\widetilde{W}_{\alpha}
\approx
\widetilde{W}_{-}+
\widetilde{W}_{+}+
\widetilde{W}_{\mathrm{osc}}
\label{eq:  WigTwidProx}
\end{equation}
where
\begin{multline*}
 \widetilde{W}_{\pm} (t,x,p,x_1,\dots,x_N)
\\
=
\mathrm{const.} \times
\exp\biggl[
          -\sum_{r=1}^{N} \frac{m_{r}\omega_{r}}{\hbar}
          \bigl(x_r - q_r(x,p)\bigr)^2
   - \tilde{\sigma}_{\pm}^{2}
          \bigl( p \mp p_{\mathrm{cl}}(x)\bigr)^2
          \biggr] \rho_{\pm} (x) 
\end{multline*}
and 
\begin{multline*}
 \widetilde{W}_{\mathrm{osc}}(t,x,p,x_1,\dots,x_N)  
\\
 =
\mathrm{const.} \times
   \exp \biggl[ 
          -\sum_{r=1}^{N} \frac{m_{r} \omega_{r}}{\hbar}
          \bigl(x_r - q_r(x,p)\bigr)^2 
          -\tilde{\sigma}_{2}^{2}
\bigl(p+\tilde{\beta} 
            p_{\mathrm{cl}}(x)\bigr)^2-\tilde{\sigma_{1}}^{2} 
            \bigl(p_{\mathrm{cl}}(x)\bigr)^2\biggr]
\\ 
\times  
            \cos \left[ \frac{2}{\hbar} S(x) + \tilde{\chi}(t,x,p)\right]
\bigl(\rho_{-}(x) \rho_{+}(x)
\bigr)^\frac{1}{2}
\end{multline*}
with
{\allowdisplaybreaks
\begin{align}
\rho_{\pm} & = \left|g_{\pm}\right|^2 \\
\tilde{\sigma}_{\pm}^{2} & = 
\frac{ \tilde{\Delta}}{\tilde{a} \pm 2 \tilde{c} p'_{\mathrm{cl}}(x)
   +\tilde{b} \bigl(p'_{\mathrm{cl}}(x)\bigr)^2}  
\label{eq:  SigPmTwid}   
\\
\tilde{\sigma}_{1}^{2} & =
\frac{\tilde{\Delta}}{\hbar^2
 \tilde{a} \tilde{\Delta}+\tilde{b}
\bigl(p'_{\mathrm{cl}}(x)\bigr)^2}
\label{eq:  Sig1Twid}
\\
\tilde{\sigma}_{2}^{2} & =
\frac{\hbar^2 \tilde{a} \tilde{\Delta} + \tilde{b} 
   \bigl(p'_{\mathrm{cl}}(x)\bigr)^2}{
    \hbar^2  \tilde{a}^2 +\left(1-2
     \hbar^2 \tilde{c}^2 + \hbar^4 \tilde{\Delta}^2\right)
     \bigl(p'_{\mathrm{cl}}(x)\bigr)^2
     +\hbar^2  \tilde{b}^2 \bigl(p'_{\mathrm{cl}}(x)\bigr)^4 }
\\
\tilde{\beta} & =
\frac{
        \tilde{c}\left(1+\hbar^2\tilde{\Delta}\right)
        p'_{\mathrm{cl}}(x) 
        }{
        \hbar^2 \tilde{a} \tilde{\Delta} + \tilde{b} 
        \bigl(p'_{\mathrm{cl}}(x)\bigr)^2
        }
\end{align}
$\tilde{\chi}(t,x,p)$ is a phase.}
 
If
$\tilde{\sigma}_{1} p_{\mathrm{cl}}\gg1$, then
$\widetilde{W}_{\mathrm{osc}}$ will be negligible and
\begin{equation*}
  \widetilde{W}_{\alpha} \approx \widetilde{W}_{-} +\widetilde{W}_{+}
\end{equation*}
For a given value of $x$ the Gaussian in the expression for
$\widetilde{W}_{\pm}$ is peaked at the point
$p=\pm p_{\mathrm{cl}}(x)$, $x_r = q_r\bigl(x,\pm p_{\mathrm{cl}}(x)\bigr)$.
Suppose that $x_r = q_r\bigl(x,p_{\mathrm{cl}}(x)\bigr)$ for
$r=1,\dots,N$.  Then
\begin{equation*}
  \widetilde{W}_{-} = \mathrm{const.} \times
  \exp\biggl[ - \tilde{\sigma}_{3}^{2}
              \bigl(p - p_{\mathrm{cl}}(x)\bigr)^2
              - \tilde{\sigma}_{-}^{2}
              \bigl(p + p_{\mathrm{cl}}(x)\bigr)^2
      \biggr]
\end{equation*}
where 
[\emph{c.f.}\ Eq.~(\ref{eq:  qrDef})]
\begin{equation}
  \tilde{\sigma}_{3}^{2} = \frac{2}{\hbar m^2}
    \int_{0}^{\infty} d \omega' \, I(\omega')
      \frac{\left(\sin \omega' t-\omega' t \cos \omega' t \right)^2
      }{{\omega'}^4}
\label{eq:  sig3TermsI}
\end{equation}
It follows that, if $\tilde{\sigma}_{-} p_{\mathrm{cl}},
\tilde{\sigma}_{3} p_{\mathrm{cl}} \gg 1$, then $\widetilde{W}_{-}$
 will be negligible, so that
 \begin{equation*}
  \widetilde{W}_{\alpha} \approx \widetilde{W}_{+}
\end{equation*}
Referring to Eq.~(\ref{eq:  vbTermsWigB}) we see that this implies
$\vb^{(\alpha)} \approx + p_{\mathrm{cl}}/m$.
Suppose, on the other hand, that
$x_r = q_r\bigl(x,-p_{\mathrm{cl}}(x)\bigr)$ for
$r=1,\dots,N$.  We then find that $\widetilde{W}_{+}$ will be
negligible if $\tilde{\sigma}_{+} p_{\mathrm{cl}},
\tilde{\sigma}_{3} p_{\mathrm{cl}} \gg 1$, in which 
case $\vb^{(\alpha)} \approx - p_{\mathrm{cl}}/m$.
The configuration space probability density function is obtained
from $\widetilde{W}_{\alpha}$ by integrating out the momentum.  It follows,
that  with probability close to 1, either $x_r \approx
q_r\bigl(x,p_{\mathrm{cl}}(x)\bigr)$ for
$r=1,\dots,N$, or $x_r \approx q_r\bigl(x,-p_{\mathrm{cl}}(x)\bigr)$ for
$r=1,\dots,N$.
 We conclude that there will be a high probability 
of $\vb^{(\alpha)}$ being close to one of the two classical values $\pm
p_{\mathrm{cl}}/m$ provided
\begin{equation}
 \tilde{\sigma}_{1} p_{\mathrm{cl}} \gg 1
\hspace{0.5 in }
\tilde{\sigma}_{\pm} p_{\mathrm{cl}} \gg 1
\hspace{0.5 in }
\tilde{\sigma}_{3} p_{\mathrm{cl}} \gg 1
\label{eq:  ClassCond}
\end{equation}

Let us now specialise to the case of the Caldeira-Leggett model, for which
the spectral density defined by Eq.~(\ref{eq:  IomegDef}) takes the 
form~\cite{Caldeira}
\begin{equation}
  I(\omega')
=\begin{cases} \frac{2 m \gamma \omega'}{\pi} \hspace{0.5 in} 
                                  & 0\le\omega'\le \Omega
              \\
              0   & \Omega < \omega'
  \end{cases}
\label{eq:  IforCaldLegg}
\end{equation}
for some cut-off frequency $\Omega$.  Subsituting this expression
in Eq.~(\ref{eq:  MTwiddleDef}) gives~\cite{Grad}
\begin{align*}
\widetilde{\mathbf{M}}(t) & \approx \frac{4\hbar m\gamma}{\pi}
             \int_{0}^{\Omega t} du\,
       \begin{pmatrix} \frac{t^2 \left(u-\sin u\right)^2}{
                      m^2 u^3} &
                     - \frac{t\left(u-\sin u\right)
                           \left(1-\cos u\right)}{m u^2}
                \\
                      -\frac{t\left(u-\sin u\right)
                           \left(1-\cos u\right)}{m u^2} &
                        \frac{\left(1-\cos u\right)^2}{u}
       \end{pmatrix}
\\
& = \frac{4\hbar m\gamma}{\pi}
       \begin{pmatrix}\frac{t^2}{m^2}
                     \left(\ln\Omega t-\frac{1}{2}-\ln 2
                            -\gamma_{\mathrm{E}}
                      \right)
       &-\frac{t}{m}\left(\ln\Omega t-\ln 2 +\gamma_{\mathrm{E}}
                    \right)
       \\-\frac{t}{m}\left(\ln\Omega t-\ln 2 +\gamma_{\mathrm{E}}
                    \right)
       & \frac{3}{2}\ln\Omega t -\frac{1}{2}\ln 2 
              +\frac{3}{2} \gamma_{\mathrm{E}}
    \end{pmatrix}
    +O\left(\tfrac{1}{\Omega t}\right)
\end{align*}
where $\gamma_{\mathrm{E}}$ is Euler's constant.
Referring to Eqs.~(\ref{eq:  DeltaTwiddleDef}) we deduce that, 
if $\ln\Omega t
\gg 1$,
\begin{equation*}
  \begin{pmatrix} \tilde{a} & \tilde{c}\\ \tilde{c} & \tilde{b}
  \end{pmatrix}
 \approx
  \frac{\pi}{2 \hbar m \gamma \ln \Omega t}
  \begin{pmatrix} \frac{3 m^2}{2 t^2} & \frac{m}{t} 
      \\ \frac{m}{t} & 1
  \end{pmatrix} 
\hspace{0.5 in} \text{and} \hspace{0.5 in}
\tilde{\Delta}  \approx
\frac{\pi^2}{8 \hbar^2  \left(\gamma t \ln \Omega t\right)^2}
\end{equation*}
It can be seen that $\tilde{a}, \tilde{b},\tilde{c}$ and $\tilde{\Delta}$
are  cut-off dependent (unlike the quantities $a,b,c$  and
$\Delta$ considered in the last
Section).  

With these values inequalities~(\ref{eq:  WigTwidProxCond}) 
 become
\begin{equation}
  (\omega t)^2
  \left(\frac{\deBrog}{\amp}\right)^{\frac{1}{3}}
  \ll 
\omega \gamma t^2 \ln \Omega t
  \ll \left(\frac{\amp}{\deBrog}\right)^{\frac{1}{3}}
\label{eq:  CLValidCond}
\end{equation}
where $\deBrog = \hbar / \left(m \omega \amp\right)$.

Eqs.~(\ref{eq:  sig3TermsI}) and~(\ref{eq:  IforCaldLegg}) imply~\cite{Grad}
\begin{equation*}
 \tilde{\sigma}_{3}^{2}
 = \frac{4 \gamma t^2}{\hbar m \pi}
  \int_{0}^{\Omega t} du \, 
        \frac{(\sin u - u \cos u)^2}{u^3}
\approx \frac{2 \gamma t^2 \ln \Omega t}{\pi
\hbar m}
\end{equation*}
if $\ln \Omega t\gg 1$ (where $\gamma_{\mathrm{E}}$ is Euler's
constant, as before).
If $x$ is not too close to one of the classical turning points
$p_{\mathrm{cl}}(x)\sim m \omega \amp$ and
${p'}_{\mathrm{cl}}(x)\sim m \omega $.  Consequently
\begin{align*}
  \tilde{\sigma}_{3}^{2}p_{\mathrm{cl}}^{2}(x) 
& \sim \frac{2 \amp}{ \pi \deBrog} \omega \gamma t^2 \ln \Omega t
\\
\intertext{and [\emph{c.f.} Eq.~(\ref{eq:  Sig1Twid})]}
\tilde{\sigma}_{1}^{2} p_{\mathrm{cl}}^{2}(x) 
& \sim
\frac{4 \amp}{3 \pi \deBrog} \frac{\omega \gamma t^2 \ln \Omega t
                       }{1+ \frac{16}{3 \pi^2}
                       \left(\omega \gamma t^2 \ln \Omega t\right)^2}
\end{align*}
Taking into account the fact that these equations assume that $t$ is in the
range specified  by inequalities~(\ref{eq:  CLValidCond}), we see
that $\tilde{\sigma}_{1} p_{\mathrm{cl}}$ and
$\tilde{\sigma}_{3} p_{\mathrm{cl}}$ will be
$ \gg 1$ provided $\omega \gamma t^2 \ln  \Omega t \gg \deBrog / \amp $.

Referring to Eq.~(\ref{eq:  SigPmTwid}) we see that,
away from the classical turning points,
\begin{equation*}
  \tilde{\sigma}_{\pm}^{2} p_{\mathrm{cl}}^{2}
\sim
  \frac{\pi \amp}{6 \deBrog} \frac{\omega}{\gamma \ln \Omega t}
  \left(1\pm \frac{4}{3}\omega t+\frac{2}{3} (\omega t)^2\right)^{-1}
\approx
\frac{\pi \amp}{6 \deBrog} \frac{\omega}{\gamma \ln \Omega t}
\end{equation*}
(since we are assuming $\omega t\ll 1$).  Consequently
$\tilde{\sigma}_{\pm} p_{\mathrm{cl}} \gg 1$ if
$\ln \Omega t \ll \omega \amp/ \left(\gamma \deBrog\right)$.  In the case
of weak coupling (so that $\gamma \ll \omega$) and large quantum numbers
(so that $\deBrog \ll \amp$) this inequality is automatically satisfied,
for all physically reasonable values of $\Omega t$.
The classicality conditions~(\ref{eq:  ClassCond}) then reduce to the single
requirement
$t \gg
\tpcrit$, where
$\tpcrit$ is the solution to
\begin{equation*}
 \omega \gamma \tpcrit^{2} \ln \Omega \tpcrit
= \frac{\deBrog}{\amp}
\end{equation*}

Finally, let us consider the relation between $\tpcrit$ and
the quantity $\tcrit$ discussed in the last section [see
Eq.~(\ref{eq:  tcDef})].  We clearly ought to have 
$\tpcrit \ge \tcrit $; however, it is not immediately apparent that this
is necessarily  the case.
In fact, the seeming difficulty disappears once it is
recalled that, in  deriving the Caldeira-Leggett master equation
from a Brownian motion model,
it is assumed~\cite{Caldeira} that $\bolt T\gg \hbar \Omega  $ and
$\Omega t \gg 1$.  In particular, the discussion in the last Section 
tacitly assumes $\Omega
\tcrit
\gg 1$. Hence
\begin{equation*}
  \omega \gamma \tcrit^2 \ln \Omega \tcrit
\ll \omega \gamma \Omega \tcrit^3
\ll \frac{\omega \gamma \bolt T \tcrit^3}{\hbar}
= \frac{3\deBrog}{2 \amp}
\end{equation*}
which implies that $\tcrit \ll \tpcrit$.
\section{Conclusion}
The motive for this investigation was the question, whether
it is true that any process which tends to produce decoherence also tends to 
make the Bohmian trajectory of a macroscopic object approximately classical.  The
results we have obtained provide some support for this hypothesis.  However, it would
clearly require  more work to settle the question.
We have only considered a particular, idealised model of the interaction between a
macroscopic body and its environment.  Moreover, our results were obtained on the
assumption that the ensemble described by $\hat{\rho}_{\mathrm{bath}}$ consists 
of coherent states [see Eq.(\ref{eq:  rhoEnvB}), and discussion in paragraphs
following].  It would clearly be desirable to see if similar results hold
in the case of other models, and for other  choices of  
ensemble\footnote{
  In this connection we should
  mention a recent paper by Geiger \emph{et al}~\cite{Geiger}, in which the authors
attempt to
  derive approximately classical Bohmian trajectories by making certain postulates
  regarding the form of the many-body wave function describing a macroscopic object.
  }.
  
However, realistic models of the interaction
between a macroscopic object and its enviroment are very complicated, so
that detailed calculations, of the kind carried out in this paper, are not usually
feasible.  What one needs is  a general principle, or mechanism, which can be shown
to be operative even in those cases where the complexity of the problem makes
detailed calculation impracticable.  The most promising candidate for such a
mechanism is the process of environmental monitoring.

Particularly relevant in this respect is a recent paper by Halliwell~\cite{HalliD}. 
Halliwell analyses Brownian motion models of the kind considered in this paper, and he
shows that the positions and momenta of the heat bath oscillators
constitute a store of information about the trajectory of the
system particle.  He also shows that there is a relationship between the amount of
information stored in the environment and the amount of decoherence.  One may
plausibly speculate that a similar principle holds true with regard to the Bohm
Interpretation:  namely, that there is a direct relationship between the degree to
which the Bohmian trajectory is approximately classical, and the amount of
information about the trajectory which is stored in the environment.
It would also be interesting to know whether such a principle applies to some of the
other interpretations which have been proposed in which the particles follow 
determinate trajectories~\cite{Nel2,Suth,Holland3,Ghir,Singh3}.
\appendix
\section{Closed Form Expressions for  $\mathbf{A}$, $\mathbf{B}_{r}$,
$\mathbf{C}_{r}$ and
$\mathbf{D}_{r r'}$}
\label{app:  ExactQBrown}
The analysis in Section~\ref{sec:  QuantBrown} is based on Halliwell
and Yu~\cite{HalliC}.  However, Halliwell and Yu 
do not give explicit expressions for
the matrices
$\mathbf{A}$,
$\mathbf{B}_r$,
$\mathbf{C}_r$ and $\mathbf{D}_{r r'}$.  The purpose of this appendix is to
derive such expressions.
Define
\begin{equation*}
  \boldsymbol{\sigma}_{0} = \begin{pmatrix} 0 & \frac{1}{m \omega_{0}} \\
                          -m \omega_{0} & 0\end{pmatrix}
\hspace{0.5 in}
  \boldsymbol{\sigma}_{r} = \begin{pmatrix} 0 & \frac{1}{m_r \omega_{r}} \\
                          -m_r \omega_{r} & 0\end{pmatrix}
\hspace{0.5 in}
  \boldsymbol{\sigma}_{-} = \begin{pmatrix} 0 & 0 \\
                         1 & 0\end{pmatrix}
\end{equation*}
Referring to Eqs.~(\ref{eq:  etaEvol})
and~(\ref{eq:  etarEvol}), and to the classical analogue of
Eq.~(\ref{eq:  BrownHamDef}), we see that
$\mathbf{A}$,
$\mathbf{B}_r$,
$\mathbf{C}_r$ and $\mathbf{D}_{r r'}$ satisfy
{\allowdisplaybreaks
\begin{align*}
 \frac{d}{d t} \mathbf{A}(t) & = \omega_0 \boldsymbol{\sigma}_{0}
 \mathbf{A}(t) -
\sum_{r=1}^{N} 
       \kappa_r \boldsymbol{\sigma}_{-} \mathbf{C}_r(t) &
 \frac{d}{d t} \mathbf{B}_r (t) & = \omega_0 \boldsymbol{\sigma}_{0}
 \mathbf{B}_r(t) -
\sum_{r'=1}^{N} 
       \kappa_{r'} \boldsymbol{\sigma}_{-}
      \mathbf{D}_{r' r}(t) \\
 \frac{d}{d t} \mathbf{C}_r (t) & = \omega_r \boldsymbol{\sigma}_r
 \mathbf{C}_r(t) - 
       \kappa_{r} \boldsymbol{\sigma}_{-} \mathbf{A}(t) &
 \frac{d}{d t} \mathbf{D}_{r r'} (t) & = \omega_r \boldsymbol{\sigma}_r 
 \mathbf{D}_{r
r'}(t) -
       \kappa_{r} \boldsymbol{\sigma}_{-} \mathbf{B}_{r'}(t) \\
\end{align*}
subject to the initial conditions $\mathbf{A}(0)=\boldsymbol{1}$,
$\mathbf{B}_{r}(0)=\mathbf{C}_{r}(0)=\boldsymbol{0}$ and
$\mathbf{D}_{r r'}(0)=\delta_{r r'} \boldsymbol{1}$.
It is convenient to re-write these equations in integral form:
\begin{align}
  \mathbf{A}(t) & = e^{\omega_{0} t \boldsymbol{\sigma}_{0}}
       - \sum_{r=1}^{N} \kappa_r
          \int_{0}^{t} dt' \,
          e^{\omega_{0} (t-t') \boldsymbol{\sigma}_{0}}
          \boldsymbol{\sigma}_{-}
          \mathbf{C}_{r}(t')
\label{eq:  AIntEq}
\\
  \mathbf{B}_{r}(t) & =
  - \sum_{r'=1}^{N} \kappa_{r'}
    \int_{0}^{t} dt' \,
    e^{\omega_{0} (t-t') \boldsymbol{\sigma}_{0}}
    \boldsymbol{\sigma}_{-} \mathbf{D}_{r' r} (t')
\label{eq:  BIntEq}
\\
  \mathbf{C}_{r}(t) & =
  - \kappa_{r}
    \int_{0}^{t} dt' \,
    e^{\omega_{r} (t-t') \boldsymbol{\sigma}_{r}}
    \boldsymbol{\sigma}_{-} \mathbf{A}(t')
\label{eq:  CIntEq}
\\
  \mathbf{D}_{r r'}(t) & =
   \delta_{r r'} e^{\omega_{r} t \boldsymbol{\sigma}_{r}}
  - \kappa_{r}
    \int_{0}^{t} dt' \,
    e^{\omega_{r} (t-t') \boldsymbol{\sigma}_{r}}
    \boldsymbol{\sigma}_{-} \mathbf{B}_{r'} (t')
\label{eq:  DIntEq}
\end{align}
Eqs.~(\ref{eq:  AIntEq}) and~(\ref{eq:  CIntEq}) imply
\begin{align}
\mathbf{A}(t) & = e^{\omega_{0} t \boldsymbol{\sigma}_{0}}
   + \int_{0}^{t} dt' \mathbf{L}(t-t') \mathbf{A}(t')
\label{eq:  AIntEqb}
\\
\intertext{while Eqs.~(\ref{eq:  BIntEq}) and~(\ref{eq:  DIntEq}) give}
\mathbf{B}_{r}(t) & = 
  - \kappa_{r} \int_{0}^{t} dt' 
   e^{\omega_{0} (t-t') \boldsymbol{\sigma}_{0}}
   \boldsymbol{\sigma}_{-}
   e^{\omega_{r} t' \boldsymbol{\sigma}_{r}}
   + \int_{0}^{t} dt'\mathbf{L}(t-t') \mathbf{B}_{r} (t')
\label{eq:  BIntEqb}
\\
\intertext{where}
\mathbf{L}(t) & = 
\sum_{r=1}^{N}  \left(\kappa_{r}^{2}
\int_{0}^{t} dt' \,
   e^{\omega_{0} (t-t') \boldsymbol{\sigma}_{0}}
   \boldsymbol{\sigma}_{-}
   e^{\omega_{r} t' \boldsymbol{\sigma}_{r}}
   \boldsymbol{\sigma}_{-} \right)
\notag
\end{align}
$\mathbf{B}_r$ can be expressed in terms of $\mathbf{A}$:}
\begin{equation}
  \mathbf{B}_{r}(t)
= - \kappa_{r} \int_{0}^{t} dt' \,
     \mathbf{A}(t') \boldsymbol{\sigma}_{-}
     e^{\omega_{r} (t-t') \boldsymbol{\sigma}_{r}}
\label{eq:  BIntEqc}
\end{equation}
as can be verified by substituting this expression
into Eq.~(\ref{eq:  BIntEqb}) and using 
Eq.~(\ref{eq:  AIntEqb}).}

Carrying out the integration in the expression for $\mathbf{L}$ we find
\begin{align*}
  \mathbf{L}(t) & =
  \begin{pmatrix} \chi (t) & 0 \\ m \dot{\chi}(t) & 0
  \end{pmatrix}
\\
\intertext{where}
  \chi (t) & =
  \frac{2}{m \omega_{0}} \int_{0}^{\infty} d\omega' \,
  I(\omega') \frac{\omega' \sin \omega_{0} t - 
                    \omega_{0} \sin \omega' t}{
                    {\omega'}^{2}-\omega_{0}^{2}}
\end{align*}
$I(\omega')$ being the spectral density 
defined by Eq.~(\ref{eq:  IomegDef}).  Let $g$ be the solution to the
integral equation
\begin{equation*}
  g(t) = \sin \omega_0 t + \int_{0}^{t} dt' \, \chi(t-t') g(t')
\end{equation*}
In terms of this function the solution to Eq.~(\ref{eq:  AIntEqb}) is
\begin{equation}
  \mathbf{A}(t) = \frac{1}{m \omega_0}
                  \begin{pmatrix} m \dot{g}(t) & g(t) \\
                                  m^2 \ddot{g}(t) & m \dot g(t)
                  \end{pmatrix}
\label{eq:  AMatExact}
\end{equation}
Eqs.~(\ref{eq:  CIntEq}), (\ref{eq:  DIntEq}) 
and~(\ref{eq:  BIntEqc}) then imply
{\allowdisplaybreaks
\begin{align}
  \mathbf{B}_{r}(t) & = \frac{ \kappa_r}{m m_r \omega_0 \omega_r}
                   \begin{pmatrix} m_r \dot{h}_{r}(t)  & h_r (t)
                             \\
                         m m_r \ddot{h}_r(t) & m\dot{h}_r(t)
           \end{pmatrix}  
\label{eq:  BMatExact}
\\
  \mathbf{C}_{r}(t) & = \frac{ \kappa_r}{m m_r \omega_0 \omega_r}
                   \begin{pmatrix} m \dot{h}_{r}(t)  & h_r (t)
                              \\
                         m m_r \ddot{h}_r(t) & m_r\dot{h}_r(t)
           \end{pmatrix}
\label{eq:  CMatExact}
\\
  \mathbf{D}_{r r'}(t) & = 
             \delta_{r r'}
             e^{\omega_r t \boldsymbol{\sigma}_r}
             +  \frac{\kappa_r \kappa_{r'}}{m m_r m_{r'}
                     \omega_0 \omega_r \omega_{r'}}
                   \begin{pmatrix}
                        m_{r'} \dot{f}_{r r'}(t)  & f_{r r'} (t) \\
                         m_{r} m_{r'} \ddot{f}_{r r'}(t) &
                          m_r\dot{f}_{r r'}(t)
           \end{pmatrix}
\label{eq:  DMatExact}
\end{align}
where
\begin{align*}
  h_{r} (t) & = - \int_{0}^{t} dt' \,  g(t-t')\sin \omega_{r}t'  \\
  f_{r r'}(t) & = \int_{0}^{t} dt' \,
                  g(t-t')
           \frac{\omega_r \sin \omega_{r'} t' - 
                  \omega_{r'} \sin \omega_{r} t'}{
                  \omega_{r}^{2} - \omega_{r'}^{2}}
\end{align*}
We see that all four matrices may be expressed in terms of the single
function $g$.}

Eqs.~(\ref{eq:  AMatExact}--\ref{eq:  DMatExact}) are exact.
Let us now consider the case of weak coupling.  Working to first order
in the $\kappa_{r}$ we have
\begin{equation*}
  g(t) = \sin \omega_{0} t + O(\kappa^2)
\end{equation*}
and consequently
{\allowdisplaybreaks
\begin{align*}
  \mathbf{A}(t) & = e^{\omega_0 t \boldsymbol{\sigma}_{0}}
  +O(\kappa^2) 
\\
  \mathbf{B}_{r}(t) & = \frac{ \kappa_r}{m m_r \omega_0 \omega_r}
                   \begin{pmatrix} m_r \dot{h}_{r}^{(0)}(t)  & 
                   h_{r}^{(0)}(t)
                         \\
                         m m_r \ddot{h}_{r}^{(0)}(t) & 
                         m\dot{h}_{r}^{(0)}(t)
           \end{pmatrix} +O(\kappa^3) 
\\
  \mathbf{C}_{r}(t) & = \frac{\kappa_r}{m m_r \omega_0 \omega_r}
                   \begin{pmatrix} m \dot{h}_{r}^{(0)}(t)  & 
                   h_{r}^{(0)} (t)
                         \\
                         m m_r \ddot{h}_{r}^{(0)}(t) & 
                         m_r\dot{h}_{r}^{(0)}(t)
           \end{pmatrix} + O(\kappa^3)\\
  \mathbf{D}_{r r'}(t) & = 
             \delta_{r r'}
             e^{\omega_r t \boldsymbol{\sigma}_r}
             +  O(\kappa^2)
\end{align*}
where}
\begin{equation*}
  h_{r}^{(0)} (t) =
  - \int_{0}^{t} dt' \,  \sin \omega_{0} (t-t')\sin \omega_{r}t'
  = -\frac{\omega_{r} \sin
\omega_{0} t  -\omega_{0}  
     \sin \omega_{r} t}{\omega_{r}^{2} - \omega_{0}^{2}}
\end{equation*}
We also note that the frequency counterterm is $O(\kappa^2)$, so 
$\omega_{0}=\omega$ to this order of approximation.  This proves
Eqs.~(\ref{eq:  AProx}--\ref{eq:  DProx}).

\section{Expression for $\mathbf{M}$ in Terms 
of  $\mathbf{A}$ and $\mathbf{B}_{r}$}
\label{app:  RedWigProp}
The purpose of this appendix is to derive the relationship between  the matrices
$\mathbf{A}$,
$\mathbf{B}_r$,
$\mathbf{C}_r$,
$\mathbf{D}_{r r'}$, and the matrix $\mathbf{M}$ which appears in
the integrated form of the Master Equation, Eq.~(\ref{eq:  RedWigProp}).  
We will also
show that the matrix $\mathbf{A}$ appearing in 
Eq.~(\ref{eq:  RedWigProp}) is the same as the matrix
$\mathbf{A}$ derived in Appendix~\ref{app:  ExactQBrown}.

We begin by noting that $\mathbf{D}_{r r'}$, regarded as a 
$2N \times 2 N$ matrix, is invertible.  In fact, it follows from
the time-reversibility of the classical equations of motion that
{\allowdisplaybreaks
\begin{align}
  \mathbf{A}(t) \mathbf{A}(-t) 
  + \sum_{r=1}^{N} \mathbf{B}_{r}(t) \mathbf{C}_{r}(-t) 
  & = \mathbf{1}
\label{eq:  RevA}
\\
  \mathbf{C}_{r}(t) \mathbf{B}_{r'}(-t) 
  + \sum_{r''=1}^{N} \mathbf{D}_{r r''}(t) \mathbf{D}_{r'' r'}(-t) 
  & = \delta_{r r'} \mathbf{1}
\label{eq:  RevB}
\\
  \mathbf{A}(t) \mathbf{B}_{r}(-t) 
  + \sum_{r'=1}^{N} \mathbf{B}_{r'}(t) \mathbf{D}_{r' r}(-t) 
  & = \mathbf{0}
\label{eq:  RevC}
\\
  \mathbf{C}_{r}(t) \mathbf{A}(-t) 
  + \sum_{r'=1}^{N} \mathbf{D}_{r r'}(t) \mathbf{C}_{r'}(-t) 
  & = \mathbf{0}
\label{eq:  RevD}
\end{align}
It is then straightforward to verify that }
\begin{equation}
  \sum_{r''=1}^{N} \mathbf{D}_{r r''} (t) 
               \mathbf{D}_{r'' r'}^{-1} (t)
= \delta_{r r'} \mathbf{1}
\label{eq:  Dmin1Prop}
\end{equation}
where
\begin{equation}
   \mathbf{D}_{r r'}^{-1} (t)
=  \mathbf{D}_{r r'}(-t) 
   - \mathbf{C}_{r}(-t) \mathbf{A}^{-1}(-t) \mathbf{B}_{r'}(-t)
\label{eq:  Dmin1Def}
\end{equation}

As before, we assume that at time $t$ system$+$environment are described by the
density matrix
\begin{equation*}
\hat{\rho}(t) =
  e^{- i t \hat{H}/\hbar}
  \bigl(
  \ket{\psi_{\mathrm{sys}}}\bra{\psi_{\mathrm{sys}}}  \otimes 
  \hat{\rho}_{\mathrm{bath}}
  \bigr)
  e^{i t \hat{H}/\hbar}
\end{equation*}
where $\hat{\rho}_{\mathrm{bath}}$ is the thermal state
defined by Eq.~(\ref{eq:  RhoBatDef}).  Let 
$W(t,\boldsymbol{\eta},\boldsymbol{\eta}_1,\dots
\boldsymbol{\eta}_N)$ be the corresponding Wigner function.
Then
\begin{equation*}
  W_{\mathrm{red}}(t,\boldsymbol{\eta})
= \int d^{2} \boldsymbol{\eta}_1 \dots d^{2} \boldsymbol{\eta}_N \,
    W_{\hat{\rho}}(t,\boldsymbol{\eta},\boldsymbol{\eta}_1,\dots,
    \boldsymbol{\eta}_N)
\end{equation*}
We have, by essentially the same argument as the one leading
to Eq.~(\ref{eq:  WtildeTermsWsys}), 
\begin{equation*}
  W_{\mathrm{red}}(t,\boldsymbol{\eta})
= \int d^{2} \boldsymbol{\eta}' \, 
  G(t,\boldsymbol{\eta}|\boldsymbol{\eta}')
  W_{\mathrm{red}}(0,\boldsymbol{\eta}')
\end{equation*}
where
\begin{align}
& G(t,\boldsymbol{\eta}|\boldsymbol{\eta}')
\notag
\\
& = \mathscr{K}
  \int d^{2} \boldsymbol{\xi} d^{2}\boldsymbol{\eta}_1 \dots d^{2} \boldsymbol{\eta}_N
\,
    \exp 
      \biggl[ - \sum_{r=1}^{N}
                  \biggl( \tanh \Bigl( \frac{\beta_r}{2}\Bigr)
                       \Bigl(\mathbf{C}_r (-t) \boldsymbol{\eta}
                     + \sum_{r'=1}^{N} \mathbf{D}_{r r'}(-t)
                       \boldsymbol{\eta}_{r'}
                        \Bigr)^{\mathrm{T}}
                  \biggr.
    \biggr.
\notag
\\
& \hspace{0.47 in} 
   \biggl.
               \biggl.
                 \boldsymbol{\Lambda}_{r}
                \Bigl(\mathbf{C}_r (-t) \boldsymbol{\eta}
                     + \sum_{r''=1}^{N} \mathbf{D}_{r r'}(-t)
                       \boldsymbol{\eta}_{r'}
                 \Bigr) 
               \biggr)
   + i \boldsymbol{\xi}^{\mathrm{T}}
      \Bigl( \boldsymbol{\eta}'-\mathbf{A}(-t) \boldsymbol{\eta}
            - \sum_{r=1}^{N} \mathbf{B}_{r}(-t) \boldsymbol{\eta}_{r}
      \Bigr)
   \biggr]
\label{eq:  RedWigPropB}
\end{align}
and where 
$\boldsymbol{\Lambda}_{r}$ is the matrix defined by
Eq.~(\ref{eq:  LambdaDef}), $\beta_{r}$ denotes the ratio $\hbar
\omega_{r}/(\bolt T)$ and $\mathscr{K}$ is a normalisation constant.

Making the substitution ${\boldsymbol{\eta}'}_{r}
=\mathbf{C}_{r}(-t)+ \sum_{r'=1}^{N} \mathbf{D}_{r r'} (-t)
\boldsymbol{\eta}_{r'}$ in the integral on the right hand side of
Eq.~(\ref{eq:  RedWigPropB}) we obtain
\begin{align*}
 G(t,\boldsymbol{\eta}|\boldsymbol{\eta}')
& = \mathrm{const.}
  \int d^{2} \boldsymbol{\xi} d^{2}{\boldsymbol{\eta}'}_1 \dots 
  d^{2}{\boldsymbol{\eta}'}_N
\\
& \hspace{0.8 in}\times \exp 
      \biggl[ - \sum_{r=1}^{N}
                  \biggl( \tanh \Bigl( \frac{\beta_r}{2}\Bigr)
                       {\boldsymbol{\eta}'}_{r}^{\mathrm{T}}
                 \boldsymbol{\Lambda}_{r}
                       {\boldsymbol{\eta}'}_{r}
                  -i \boldsymbol{\xi}^{\mathrm{T}}
                  \mathbf{A}^{-1}(t)\mathbf{B}_{r}(t)
            {\boldsymbol{\eta}'}_{r}
                  \biggr)
      \biggr.
    \\
&  \hspace{3 in} \biggl.
     + i
\boldsymbol{\xi}^{\mathrm{T}}
      \Bigl( \boldsymbol{\eta}'-\mathbf{A}^{-1}(t) \boldsymbol{\eta} 
      \Bigr)
   \biggr]
\end{align*}
where we have used Eqs.~(\ref{eq:  Dmin1Prop}) and~(\ref{eq:  Dmin1Def}),
together with the relations [which are easily seen to follow from
Eqs.~(\ref{eq:  RevA}--\ref{eq:  Dmin1Def})]
\begin{align*}
   \sum_{r' =1}^{N} \mathbf{B}_{r'}(-t) \mathbf{D}_{r' r}^{-1}(-t)
& = -\mathbf{A}^{-1}(t) \mathbf{B}_{r}(t)
\\
  \mathbf{A}(-t) - \sum_{r,r'=1}^{N}
  \mathbf{B}_{r'}(-t) \mathbf{D}_{r' r}^{-1} (-t) \mathbf{C}_{r}(-t)
& = \mathbf{A}^{-1}(t)
\end{align*}
Carrying out the Gaussian integrations we deduce
\begin{equation*}
  G(t,\boldsymbol{\eta}|\boldsymbol{\eta}')
=  \frac{1}{\pi \det \mathbf{A} \sqrt{\det \mathbf{M}}} 
  \exp\biggl[ - \Bigl(\boldsymbol{\eta}'-\mathbf{A}^{-1}(t) 
                     \boldsymbol{\eta}
                 \Bigr)^{\mathrm{T}}
                 \mathbf{M}^{-1}(t)
                 \Bigl(\boldsymbol{\eta}'-\mathbf{A}^{-1}(t) 
                     \boldsymbol{\eta}
                 \Bigr)
     \biggr]
\end{equation*}
where
\begin{equation}
  \mathbf{M}(t) =
     \mathbf{A}^{-1}(t) \left(
\sum_{r=1}^{N} \coth \Bigl( \frac{\beta_r}{2} \Bigr)
\mathbf{B}_{r}(t)
     \boldsymbol{\Lambda}^{-1}_{r}
     \bigl( \mathbf{B}_{r}(t)
     \bigr)^{\mathrm{T}}\right) 
     \bigl(\mathbf{A}^{-1}(t)\bigr)^{\mathrm{T}}
\label{eq:  MFormulaB}
\end{equation}
and where the normalisation constant is fixed by the requirement
\begin{equation*}
  \int d\boldsymbol{\eta} \, W_{\mathrm{red}}(t, \boldsymbol{\eta}) =1
\end{equation*}
Comparing with Eq.~(\ref{eq:  RedWigProp})
we see that the matrix $\mathbf{M}$ appearing in
Eq.~(\ref{eq:  RedWigProp}) is the same as the matrix given by
Eq.~(\ref{eq:  MFormulaB}).


\begin{thebibliography}{99}
\bibitem{Bohm2}
 D.~Bohm and B.J.~Hiley,
  \emph{The Undivided Universe} (Routledge, London, 1993).
\bibitem{Holland1}
  P.R.~Holland,
  \emph{The Quantum Theory of Motion} (Cambridge University Press, Cambridge,
1993).
\bibitem{Holland2}
  P.R.~Holland,
  in \emph{Bohmian Mechanics and Quantum Theory:  An Appraisal},
  J.T.~Cushing, A.~Fine and S.~Goldstein, eds.\ (Kluwer, Dordrecht, 1996).
\bibitem{self1}
  D.M.~Appleby, Los Alamos e-print, xxx.lanl.gov,
quant-ph/9905003.
\bibitem{Griff}
  R.B.~Griffiths,
    \emph{J.~Stat.~Phys.}\ \textbf{36}, 219 (1984).
\bibitem{GellMann2}
  M.~Gell-Mann and J.B.~Hartle,
    in W.H.~Zurek (ed), \emph{Complexity, Entropy and the
Physics of Information} (Addison-Wesley, Reading, 1990).
\bibitem{GellMann1}
  M.~Gell-Mann and J.B.~Hartle,
    \emph{Phys.\ Rev. D} \textbf{47}, 3345 (1993).
\bibitem{Omnes}
 R.~Omnes,
   \emph{The Interpretation of Quantum Mechanics}
   (Princeton University Press, Princeton, 1994).
\bibitem{Zurek2}
  W.H.~Zurek,
  \emph{Phys.\ Today} \textbf{40}, October p.36 (1991).
\bibitem{Zurek1}
  W.H.~Zurek,
  \emph{Prog.\ Theor.\ Phys.\ }\textbf{89}, 281 (1993).
\bibitem{Zurek3}
  W.H.~Zurek,
  \emph{Phil.\ Trans.\ Roy.\ Soc.\ Lond.\ A} \textbf{356}, 1793 (1998).
\bibitem{Joos}
  E.~Joos and H.D.~Zeh,
 \emph{Z.~Phys.~B\ }\textbf{59}, 223 (1985).
\bibitem{Giulini}
  D.~Giulini, E.~Joos, C.~Kiefer, J.~Kupsch, 
  I.-O.~Stamatescu and H.D.~Zeh,
  \emph{Decoherence and the Appearance of a Classical World in
Quantum Theory} (Springer, Berlin, 1996).
\bibitem{Dickson}
  W.M.~Dickson,
  \emph{Quantum Chance and non-Locality}
  (Cambridge University Press, Cambridge, 1998).
\bibitem{Zeh}
 H.D.~Zeh,
   Los Alamos e-print, xxx.lanl.gov, quant-ph/9812059.
\bibitem{Caldeira}
  A.O.~Caldeira and A.J.~Leggett,
   \emph{Physica A} \textbf{121}, 587 (1983).
\bibitem{Grab}
 H.~Grabert, P.~Schramm and G-L.~Ingold,
 \emph{Phys.\ Rep.\ }\textbf{168}, 115 (1988).
\bibitem{Hu}
  B.L.~Hu, J.P.~Paz and Y.~Zhang,
\emph{Phys.\ Rev.\ D} \textbf{45}, 2843 (1992).
\bibitem{Hu2}
  B.L.~Hu, J.P.~Paz and Y.~Zhang,
\emph{Phys.\ Rev.\ D} 
  \textbf{47}, 1576 (1993).
\bibitem{Hugh}
 L.P.~Hughston, R.~Josza and W.K.~Wootters,
 \emph{Phys.\ Lett.\ A} \textbf{183}, 14 (1993).
\bibitem{HalliB}
  J.J.~Halliwell and A.~Zoupas,
  \emph{Phys.\ Rev.\ D} \textbf{55}, 4697 (1997).
\bibitem{HalliC}
 J.J.~Halliwell and T.~Yu,
 \emph{Phys.\ Rev.\ D} \textbf{53}, 2012 (1996).
\bibitem{AnglinB}
 J.~Anglin and S.~Habib,
 \emph{Mod.\ Phys.\ Lett.\ A} \textbf{11}, 2655 (1996).
\bibitem{Diosi1}
  L.~D\'{i}osi,
  \emph{Europhys.\ Lett.\ }\textbf{22}, 1 (1993).
\bibitem{Diosi2}
 L.~D\'{i}osi,
 \emph{Physica A} \textbf{199}, 517 (1993).
\bibitem{Tegmark}
 M.~Tegmark,
  \emph{Found.\ Phys.\ Lett.\ }\textbf{6}, 571 (1993).
\bibitem{Kupsch}
  J.~Kupsch,
  Los Alamos e-print, xxx.lanl.gov, quant-ph/9811010.
\bibitem{Lindblad}
 G.~Lindblad,
 \emph{Commun.\ Math.\ Phys.\ }\textbf{48}, 119 (1976).
\bibitem{Chandra}
  S.~Chandrasekhar,
  \emph{Rev.\ Mod.\ Phys.\ }\textbf{15}, 1 (1943).
\bibitem{HalliA}
  C.~Anastopoulos and J.J.~Halliwell,
  \emph{Phys.\ Rev. D} \textbf{51}, 6870 (1995).
\bibitem{Hus}
 K.~Husimi,
 \emph{Proc.\ Phys.\ Math.\ Soc.\ Jpn.\ }\textbf{22}, 264
(1940).
\bibitem{Hil}
 M.~Hillery, R.F.~O'Connell, M.O.~Scully and E.P.~Wigner,
 \emph{Phys.\ Rep.\ }\textbf{106}, 121 (1984).
\bibitem{Lee}
 H.W.~Lee,
  \emph{Phys.\ Rep.\ }\textbf{259}, 147 (1995).
\bibitem{Leon}
 U.~Leonhardt,
 \emph{Measuring the Quantum State of Light}
 (Cambridge University Press, Cambridge, 1997).
\bibitem{Glauber}
  R.J.~Glauber,
  \emph{Phys.\ Rev.\ Lett.\ }\textbf{10}, 84 (1963).
\bibitem{Sudar}
  E.C.G.~Sudarshan,
  \emph{Phys.\ Rev.\ Lett.\ }\textbf{10}, 277 (1963).
\bibitem{Grad}
  I.S.~Gradshteyn and I.M.~Ryzhik,
  \emph{Table of Integrals, Series and Products}
  (Academic Press, New York, 1980).
\bibitem{Geiger}
  H.~Geiger, G.~Obermair and Ch.~Helm,
  Los Alamos e-print, xxx.lanl.gov, quant-ph/9906082.
\bibitem{HalliD}
 J.J.~Halliwell, 
  Los Alamos e-print, xxx.lanl.gov, quant-ph/9902008.
\bibitem{Nel2}
  E.~Nelson,
    \emph{Quantum Fluctuations}
      (Princeton University Press, Princeton N.J., 1985).
\bibitem{Suth}
  R.I.~Sutherland,
    \emph{Found.\ Phys.\ }\textbf{27}, 845 (1997).
\bibitem{Holland3}
  P.R.~Holland,
   \emph{Found.\ Phys.\ }\textbf{28}, 881 (1998).
\bibitem{Ghir}
  E.~Deotto and G.C.~Ghirardi,
     \emph{Found.\ Phys.\ }\textbf{28}, 1 (1998).
\bibitem{Singh3}
 S.M.~Roy and V.~Singh,
 Los Alamos e-print, quant-ph/9811041.
\label{sec:  bibliography}
\end{thebibliography}
\end{document}